\documentstyle[prb,aps,eqsecnum,multicol,epsf]{revtex}

\newcommand{\bleq}{\ifpreprintsty
                   \else
                   \end{multicols}\vspace*{-3.5ex}{\tiny
                   \noindent\begin{tabular}[t]{c|}
                   \parbox{0.493\hsize}{~} \\ \hline \end{tabular}}
                   \fi} 
\newcommand{\eleq}{\ifpreprintsty
                  \else
                   {\tiny\hspace*{\fill}\begin{tabular}[t]{|c}\hline
                    \parbox{0.49\hsize}{~} \\
                    \end{tabular}}\vspace*{-2.5ex}\begin{multicols}{2}
                    \fi}
\newcommand{\bcols}{\ifpreprintsty\else\begin{multicols}{2}\fi}
\newcommand{\ecols}{\ifpreprintsty\else\end{multicols}\fi}

\begin{document}
\draft

\title{A new approach to strongly correlated fermion systems: \\
the spin-particle-hole coherent-state path integral }

\author{N. Dupuis }
\address{ Laboratoire de Physique des Solides, Associ\'e au CNRS,
Universit\'e Paris-Sud, 91405 Orsay, France } 
 
\date{May 3, 2001}
\maketitle

\begin{abstract}
We describe a new path integral approach to strongly correlated
fermion systems, considering the Hubbard model as a specific example. 
Our approach is based on the introduction of spin-particle-hole
coherent states which generalize the spin-$\frac{1}{2}$ coherent
states by allowing the creation of a hole or an
additional particle.  The action of the fermion system
$S[\gamma^*,\gamma;{\bf\Omega}]$ can be expressed as a function of two
Grassmann variables ($\gamma_\uparrow$,$\gamma_\downarrow$) describing
particles propagating in the lower and upper Hubbard bands, and a unit
vector field ${\bf\Omega}$ whose dynamics arises from spin
fluctuations. In the strong correlation limit,
$S[\gamma^*,\gamma;{\bf\Omega}]$ can be
truncated to quartic order in the fermionic fields and used as the
starting point of a strong-coupling diagrammatic  
expansion in $t/U$ ($t$ being the intersite hopping
amplitude and $U$ the on-site Coulomb repulsion). We discuss possible
applications of this formalism and its connection to the $t$-$J$ model
and the spin-fermion model.   
\end{abstract}


\bcols

\section{Introduction}

Understanding the properties of strongly correlated fermion systems
remains one of the main goals of condensed-matter physics. In 
narrow-band electron systems, the interplay between  strong Coulomb repulsion
(which tends to localize the electrons) and band structure effects
(which favor their itinerant character) leads to a variety of
different behaviors ranging from metallic Fermi liquid to Mott-Hubbard
insulator. Standard weak-coupling approaches fail to describe
these phenomena so that no general theoretical method is available to
analyze strongly correlated fermion systems. 
Even for the Hubbard model, \cite{Hubbard63,Gutzwiller63,Kanamori63,Gebhard97}
which is supposed to be one of
the simplest (realistic) models of strongly correlated fermions, exact
solutions or well-controlled approximations exist only in a few
special cases, like in one-dimension\cite{Voit} or in the limit of infinite
dimensions.\cite{Georges96} We describe in this paper a new approach
to strongly correlated systems, with the (two-dimensional) Hubbard
model as primary example.  

Various weak-coupling theories have  been applied to the Hubbard model
and its extensions. The simplest of these approaches, the Hartree-Fock
theory, supplemented by the random-phase approximation (RPA) for the
calculation of susceptibilities, correctly predicts an
antiferromagnetic (AF) ground-state at half-filling in both limits of
large and small on-site Coulomb repulsion $U$. \cite{Schrieffer89} 
However, it also
predicts long-range AF order at finite temperature in a
two-dimensional (2D) system, in contradiction with Mermin-Wagner
theorem.\cite{Mermin66}  Many attempts to improve on
Hartree-Fock/RPA theory can be found in the literature: paramagnon-like
theories,\cite{Kampf90,SF} fluctuation exchange approximation
(FLEX),\cite{Bickers89,Bickers91a} pseudo-potential Parquet
approach,\cite{Bickers91a,Bickers91b} 
two-particle self-consistent theory. \cite{Vilk97} Most of these approaches
meet with serious difficulties regarding a correct description of the
physical properties of the 2D Hubbard model: absence of long-range order
at finite temperature, exponential divergence of the magnetic
correlation length in $1/T$ at low temperature, existence of Hubbard
bands and/or precursors of the AF bands in the density of states,
pseudo-gap at low energy, etc. Besides, irrespective of their success
at weak Coulomb repulsion, it seems that weak-coupling
theories are doomed to fail in the strong-coupling regime when the
Coulomb repulsion exceeds the bandwidth ($U\sim 4Dt$ with
$t$ the intersite hopping amplitude and $D$ the dimensionality). [See
Ref.~\onlinecite{Vilk97} for a detailed discussion of weak-coupling
approaches to the Hubbard model.] 
As emphasized early on by Mott,\cite{Mott} a characteristic property
of strongly correlated systems is the existence of local moments
already in the metallic phase. \cite{Gebhard97} Weak-coupling theories
fail to properly describe these local moments. For instance, within
Hartree-Fock theory, local moments are totally absent in the metallic
phase and appear only in presence of long-range magnetic order. 

On the other hand, the perturbation expansion around the atomic limit,
\cite{Hubbard66,Bulaevskii73,Hone73,Plischke74,Kubo80,Zhao87,Pan91,Bartkowiak92,Izyumov92} 
which is expected to provide a reliable starting point in the strong
correlation limit, also presents its own difficulties in spite of
recent progress.\cite{Pairault98,Metzner91}  It is not completely
clear how to handle the degeneracy of the ground-state in the atomic
limit. \cite{Pairault98} Moreover, the 
expansion involves two dimensionless parameters, $t/U$ and $t/T$, and
therefore breaks down when the temperature is much below the
electronic bandwidth. 

Nonetheless, it is possible to derive an effective Hamiltonian to a
given order in $t/U$. \cite{Anderson59,Bulaevskii67,Harris67} 
At half-filling, the Hubbard model reduces to the Heisenberg
model in the strong-coupling limit. This model describes local spins
(i.e. local moments) coupled by short-range exchange interactions.
Away from half-filling, the effective
Hamiltonian in the strong correlation limit is given by the $t$-$J$
model. So far, no satisfying description of a doped Mott insulator
has emerged from the $t$-$J$ model. The Schwinger-boson slave-fermion
mean-field theory, which provides very good results at
half-filling,\cite{Arovas88}  has not been as successful in the
doped case.\cite{Jayaprakash89,note0.0}  On the other hand, the
slave-boson mean-field theory\cite{Kotliar} does not even reproduce the known
results at half-filling, so that its predictions (near half-filling) are highly
questionable. 

A very appealing path integral formulation of the $t$-$J$ model, based on the
spin-hole coherent-states $|{\bf\Omega},\zeta\rangle$, has been introduced by
Auerbach and Larson.\cite{Auerbach91} The existence of local moments
is obvious in this formulation, since a singly occupied site ${\bf r}$
is described by a spin-coherent state $|{\bf\Omega}_{\bf
r}\rangle$. The set of Grassmann variables $\zeta=\lbrace\zeta_{\bf
r}\rbrace$ allows for the presence of holes. The spin-hole
coherent-state path integral was used to study spin polarons in the
(large-$S$) semiclassical limit of the $t$-$J$ model. 

Schulz has derived a path integral formulation of the Hubbard model
which also exemplifies the existence of local moments.\cite{Schulz90,Weng91}
The Coulomb repulsion is treated within a large-$U$ Hartree-Fock
approximation, whereas the SU(2) spin-rotation symmetry is
maintained by introducing a fluctuating spin-quantization axis in the
functional integral. The resulting effective action
$S[\gamma^*,\gamma;{\bf\Omega}]$ is expressed in
terms of two Grassmann variables
($\gamma_\uparrow$,$\gamma_\downarrow$) which describe particles propagating
in the lower (LHB) and upper (UHB) Hubbard bands, and a
unit vector field ${\bf\Omega}$.\cite{note0} A singly occupied site
${\bf r}$ corresponds to a local moment pointing in the direction of
${\bf\Omega}_{\bf r}$. Similar ideas, in view of Monte Carlo
simulation, have been recently discussed by Bickers and
Scalapino.\cite{Bickers00}  

In a previous publication, Pairault and the present author have
reported a systematic $t/U$ expansion of the Hubbard model.\cite{ND}
As in Schulz's work, central to this approach is the introduction of a
fluctuating spin-quantization axis. By adapting the method of
Refs.~\onlinecite{Sarker88,Pairault98}, the $t/U$ expansion was
generated by a Grassmannian Hubbard-Stratonovich transformation of the
hopping term. As in Ref.~\onlinecite{Schulz90}, the effective action
$S[\gamma^*,\gamma;{\bf\Omega}]$ of the Hubbard model in
the strong-coupling regime is expressed as a 
function of two Grassmann variables and a unit vector field. We found however 
that the Hartree-Fock treatment of the Coulomb repulsion is not
sufficient in the strong correlation limit, since it misses processes
of order $t/U$. In particular, it does not allow to recover the
$t$-$J$ model when the UHB is integrated out.\cite{ND} 

The aim of this paper is to further develop the strong-coupling
expansion introduced in Ref.~\onlinecite{ND}. (i) The functional
integral formulation is derived from a completely different
perspective, starting from spin-particle-hole coherent states
(Sec.~\ref{sec:sph}). The latter 
generalize the spin-$\frac{1}{2}$ coherent states by allowing the
creation of a hole or the introduction of an additional particle. This
new derivation emphasizes the connection with the spin coherent-state path
integral formulation of the Heisenberg model. (ii) As in
Ref.~\onlinecite{ND}, the strong-coupling expansion is generated by
two successive Hubbard-Stratonovich transformations of the intersite
hopping term. The resulting action $S[\gamma^*,\gamma;{\bf\Omega}]$ is
obtained to {\it all} orders in $t/U$ (Sec.~\ref{sec:hs}). This action
contains interaction terms to all orders which are determined by the
(exact) atomic vertices. In the strong
correlation limit, $S[\gamma^*,\gamma;{\bf\Omega}]$ can be truncated
to quartic order in 
the fermionic fields (Sec.~\ref{sec:ea}). The action is then entirely
determined by the single-particle atomic Green's function and the
two-particle atomic vertex. (iii) We discuss in detail the
strong-coupling perturbative expansion and briefly point out
possible applications of our formalism which are developed in detail
in separate publications\cite{ND1,ND2}  (Sec.~\ref{sec:dpt}). Finally,
we discuss the connection of this formalism to the $t$-$J$ model and
the spin-fermion model.

\section{Spin-particle-hole coherent-state path integral}
\label{sec:sph}

In this section, we define the spin-particle-hole coherent states and
derive a path integral representation of the partition function. To be
specific, we consider the Hubbard model defined by
\begin{equation}
\hat H=-t \sum_{\langle{\bf r},{\bf r}'\rangle,\sigma} (\hat c^\dagger_{{\bf r}
\sigma}\hat c_{{\bf r}'\sigma} +{\rm h.c.} ) + U \sum_{\bf r} \hat n_{{\bf r}
\uparrow} \hat n_{{\bf r}\downarrow} ,
\label{Ham}
\end{equation}
where $\hat c_{{\bf r}\sigma}$ is a fermionic operator for a $\sigma$-spin 
particle at site $\bf r$ ($\sigma=\uparrow,\downarrow$), $\hat n_{{\bf
r}\sigma}=\hat c^\dagger_{{\bf r} \sigma}\hat c_{{\bf r}\sigma}$, and
$\langle {\bf r},{\bf r}'\rangle$ denotes nearest neighbors. For
simplicity, we consider a bipartite $D$-dimensional lattice. We denote
by $\mu$ the chemical potential and consider only hole doping
(i.e. $\mu \leq U/2$). We take $\hbar=k_B=1$ throughout the paper.

\subsection{Spin-particle-hole coherent-states}

\subsubsection{Definition}

We first consider a single site. A basis of the Hilbert space is
${\cal B}=\lbrace\vert0\rangle, \vert\uparrow\rangle,
\vert\downarrow\rangle, \vert\uparrow\downarrow\rangle \rbrace$, where
$\vert0\rangle$, $\vert\uparrow\rangle=\hat c_\uparrow^\dagger\vert0\rangle$,
$\vert\downarrow\rangle =\hat c_\downarrow^\dagger\vert0\rangle$ and
$\vert\uparrow\downarrow\rangle=\hat c_\uparrow^\dagger
\hat c_\downarrow^\dagger\vert0\rangle$ are the empty, 
singly occupied (with spin up and spin down) and doubly occupied
states, respectively. Instead of using the states
$\vert\uparrow\rangle$ and $\vert\downarrow\rangle$ to describe a
singly occupied site, we introduce the spin-$\frac{1}{2}$ coherent state
\begin{eqnarray}   
\vert{\bf \Omega}\rangle &=& e^{-\frac{i}{2}\varphi \sigma_z}
         e^{-\frac{i}{2}\theta \sigma_y}
         e^{-\frac{i}{2}\psi \sigma_z} \vert\uparrow\rangle \nonumber \\ 
     &=& \cos\frac{\theta}{2}e^{-\frac{i}{2}(\varphi+\psi)} \vert
         \uparrow\rangle  +
         \sin\frac{\theta}{2}e^{\frac{i}{2}(\varphi-\psi)} \vert
         \downarrow\rangle  , 
\end{eqnarray}
where $\theta$ and $\varphi$ are the polar angles determining the direction
of the unit vector ${\bf\Omega}(\theta,\varphi)$. The choice of $\psi$
is free and corresponds to a ``gauge'' freedom.\cite{Auerbach}
${\cal B}'_{\bf\Omega}=\lbrace \vert0\rangle,\vert{\bf\Omega}\rangle,
\vert\uparrow\downarrow\rangle \rbrace$ will be
used to construct the 
spin-particle-hole coherent states. Although it is not (for fixed
${\bf\Omega}$) a basis of the Hilbert space, we will show how the
whole Hilbert space can be described by varying ${\bf\Omega}$.  

We now enlarge the physical Hilbert space by introducing three bosonic
operators ($\hat e$, $\hat p$ and $\hat d$) and rewrite the states of ${\cal
B}'_{\bf\Omega}$ as 
\begin{eqnarray}
\vert0\rangle &=& \hat e^\dagger \vert {\rm vac}\rangle =
\hat\gamma_\uparrow \vert{\bf \Omega}\rangle ,  \nonumber  \\
\vert{\bf \Omega}\rangle &=& \hat p^\dagger \hat f_\uparrow^\dagger \vert {\rm
vac}\rangle , \nonumber \\ 
\vert\uparrow\downarrow\rangle &=& \hat d^\dagger \hat f_\uparrow^\dagger
\hat f_\downarrow^\dagger \vert {\rm vac}\rangle  
      = -\hat\gamma^\dagger_\downarrow \vert{\bf \Omega}\rangle ,
\end{eqnarray}
where $|{\rm vac}\rangle$ denotes the vacuum of the enlarged Hilbert
space. The latter contains
unphysical states which can be eliminated by imposing the constraints
\begin{eqnarray}
\hat Q^{(1)} &=& \hat e^\dagger\hat e + \hat p^\dagger\hat p + \hat
d^\dagger\hat d -1 = 0 , \nonumber \\ 
\hat Q^{(2)}_\uparrow &=& \hat f^\dagger_\uparrow\hat f_\uparrow -
\hat p^\dagger\hat p - \hat d^\dagger\hat d = 0 , 
\nonumber \\ 
\hat Q^{(2)}_\downarrow &=& \hat f^\dagger_\downarrow\hat f_\downarrow
- \hat d^\dagger\hat d = 0  .
\label{cons}
\end{eqnarray}
The (slave) bosonic particles are similar to those introduced
by Kotliar and Ruckenstein\cite{Kotliar86} although here the
$\hat p$ boson is spinless. Once we have introduced the
spin-$\frac{1}{2}$ coherent state $|{\bf\Omega}\rangle$, we need two
operators, $\hat\gamma_\uparrow$ and $\hat\gamma_\downarrow$, to allow
for the creation of a hole or an additional particle. These two
operators are Hubbard operators\cite{Hubbard65} ($\hat\gamma_\uparrow
\equiv |0\rangle\langle {\bf\Omega}|$ and $\hat\gamma_\downarrow
\equiv -|{\bf\Omega}\rangle\langle \uparrow\downarrow|$), and have the
following expression, 
\begin{eqnarray}
\hat\gamma_\uparrow &=& \hat e^\dagger \hat p \hat f_\uparrow , \nonumber \\
\hat\gamma_\downarrow &=& \hat p^\dagger \hat d \hat f_\downarrow ,
\label{gamdef}
\end{eqnarray}
in terms of the slave bosons and the fermionic operators $\hat
f_\sigma$. Note that 
all the operators are defined with respect to the spin-quantization
axis ${\bf\Omega}$, so that a $\hat f_\uparrow $- (or
$\hat\gamma_\uparrow$-) particle 
has its spin pointing along 
${\bf\Omega}$. Thus, the LHB is populated only with up-spin
particles (in the spin reference frame defined by ${\bf\Omega}$). This also
implies that only down-spin particles can be introduced in the UHB.
The operators $\hat\gamma_\sigma$  play a fundamental role in
our approach. Although they appear at this stage as composite
operators [see Eqs.~(\ref{gamdef})], we shall show in Sec.~\ref{sec:sce} how
the partition  
function can be written as a functional integral expressed in terms of
the {\it elementary} fields $\gamma_\sigma$ and the unit vector field
${\bf\Omega}$. 

The spin-particle-hole coherent states are defined by 
\begin{equation}
\vert {\bf \Omega},\zeta\rangle =
\exp(e\hat e^\dagger+p\hat p^\dagger+d\hat d^\dagger-f_\uparrow\hat
f^\dagger_\uparrow -f_\downarrow\hat
f^\dagger_\downarrow ) \vert
{\rm vac}\rangle ,  
\label{coh}
\end{equation}
where $f_\uparrow,f_\downarrow$ are Grassmann variables and $e,p,d$
$c$-numbers. We use the short-hand notation $\zeta\equiv
(f_\uparrow,f_\downarrow,e,p,d)$.  
By using the constraints (\ref{cons}), we can rewrite the
spin-particle-hole coherent states as
\begin{equation}
\vert {\bf \Omega},\zeta\rangle = e\vert0\rangle -pf_\uparrow\vert{\bf
\Omega}\rangle-df_\uparrow f_\downarrow \vert\uparrow\downarrow\rangle . 
\label{coh1}
\end{equation}
Eqs.~(\ref{coh}) and (\ref{coh1}) extend the definition of the
spin-$\frac{1}{2}$ coherent states by allowing for the presence of a
hole or an additional particle.

\subsubsection{Scalar product}

We consider the scalar product between the states
$\vert {\bf \Omega},\zeta\rangle$ and $\vert {\bf
\Omega}',\zeta'\rangle$. If ${\bf\Omega}={\bf\Omega}'$ all fermionic
variables are defined with respect to the same quantization axis. We
then have the standard result\cite{Negele} 
\begin{eqnarray}
\langle {\bf \Omega},\zeta\vert{\bf \Omega}',\zeta'\rangle &=& \exp
(\zeta^* \zeta') \nonumber \\ 
&\equiv & \exp (e^*e'+p^*p'+d^*d'+\sum_\sigma f^*_\sigma f_\sigma').
\label{scalar1}
\end{eqnarray} 
When ${\bf\Omega}\neq{\bf\Omega}'$, we use Eq.~(\ref{coh1}) to obtain
\begin{eqnarray}
\langle {\bf \Omega},\zeta\vert{\bf \Omega}',\zeta'\rangle &=&
e^*e'+p^*f_\uparrow^*f_\uparrow'p' +
d^*d'f_\downarrow^*f_\uparrow^*f_\uparrow'f_\downarrow' \nonumber \\ &&   
+ p^*f_\uparrow^*f_\uparrow'p' (\langle {\bf \Omega}\vert{\bf
\Omega}'\rangle -1) . 
\label{scalar2.0}
\end{eqnarray}
For ${\bf\Omega}={\bf\Omega}'$, this expression should be equivalent
to Eq.~(\ref{scalar1}), so that
\begin{equation}
\exp(\zeta^*\zeta') \equiv e^*e' +p^*f_\uparrow^*f_\uparrow'p'+
d^*d'f_\downarrow^*f_\uparrow^*f_\uparrow'f_\downarrow'        
\end{equation}
provided the constraints (\ref{cons}) are satisfied. Similarly, by
keeping only the terms which are compatible with the constraints, we
obtain
\begin{eqnarray}
\exp\bigl(\zeta^*\zeta'+f_\uparrow^*p^*f_\uparrow'p' (\langle {\bf
\Omega}\vert{\bf 
\Omega}'\rangle-1)\bigr)= e^*e' && \nonumber \\  
+p^*f_\uparrow^*f_\uparrow'p' \langle {\bf \Omega}\vert{\bf \Omega}'\rangle +
d^*d'f_\downarrow^*f_\uparrow^*f_\uparrow'f_\downarrow', &&  
\label{scalar2.1}
\end{eqnarray}
a result valid whatever the value of ${\bf\Omega}$ and ${\bf\Omega}'$. 
Comparing Eqs.~(\ref{scalar2.0}) and (\ref{scalar2.1}), we arrive at the
following expression for the scalar product of two 
spin-particle-hole coherent states in the {\it physical} Hilbert space:
\begin{equation}
\langle {\bf \Omega},\zeta\vert{\bf \Omega}',\zeta'\rangle =
\exp\bigl(\zeta^*\zeta'+f_\uparrow^*p^*f_\uparrow'p'(\langle {\bf
\Omega}\vert{\bf 
\Omega}'\rangle-1)\bigr) . 
\label{scalar2}
\end{equation}

\subsubsection{Resolution of the identity}

We seek a resolution of the identity in the form
\begin{equation}
{\cal N} \int \frac{d{\bf\Omega}}{4\pi} \int d\zeta^*d\zeta
e^{-\alpha|\zeta|^2}\vert {\bf \Omega},\zeta\rangle\langle {\bf
\Omega},\zeta \vert \hat P = \hat I,
\label{clos}
\end{equation}
where $\hat I$ is the unit operator and
\begin{equation}
\alpha |\zeta|^2 \equiv \alpha_e|e|^2+|p|^2+\alpha_d|d|^2+\sum_\sigma f^*_\sigma
f_\sigma .
\end{equation}
$\cal N$, $\alpha_e$ and $\alpha_d$ are constants to be determined in
order for Eq.~(\ref{clos}) to be satisfied. The projection operator 
\begin{equation}
\hat P = \delta_{\hat Q^{(1)},0} \prod_\sigma \delta_{\hat Q^{(2)}_\sigma,0}
\label{proj}
\end{equation}
ensures that the resolution of the identity acts only in the physical Hilbert
space. The measure in Eq.~(\ref{clos}) is defined by
\begin{equation}
d\zeta^*d\zeta = \frac{de^*de}{2i\pi}\frac{dp^*dp}{2i\pi}\frac{dd^*dd}{2i\pi}
df_\uparrow^*df_\uparrow df_\downarrow^*df_\downarrow ,
\end{equation}
where
\begin{equation}
\frac{dz^*dz}{2i\pi}=\frac{1}{\pi} d{\rm Re}(z)d{\rm Im}(z) 
\end{equation}
for a complex variable $z$. Integrating over $\zeta$, we rewrite
Eq.~(\ref{clos}) as
\begin{equation}
{\cal N} \int \frac{d{\bf \Omega}}{4\pi} \Biggl( \frac{\vert0\rangle
\langle 0\vert }{\alpha_e^2 \alpha_d} + \frac{\vert{\bf \Omega}\rangle 
\langle{\bf \Omega}\vert}{\alpha_e \alpha_d} +
\frac{\vert\uparrow\downarrow\rangle 
\langle\uparrow\downarrow\vert}{\alpha_e \alpha_d^2} \Biggr)=\hat I . 
\end{equation}
Since the spin coherent states satisfy $2(4\pi)^{-1}\int
d{\bf\Omega} \vert{\bf \Omega}\rangle \langle{\bf \Omega}\vert=
\vert\uparrow\rangle\langle\uparrow\vert
+\vert\downarrow\rangle\langle\downarrow\vert$  
(resolution of the identity in the space of singly occupied states),
we conclude that Eq.~(\ref{clos}) is satisfied for
\begin{eqnarray}
{\cal N} &=& 8 , \nonumber \\
\alpha_e &=& \alpha_d=2 . 
\end{eqnarray}
The constants $\alpha_e=\alpha_d$ ($\neq 1$) are necessary to avoid an
over-counting of the empty and doubly occupied states: for a given
${\bf\Omega}$, ${\cal B}_{\bf\Omega}'$ already contains the states
$\vert0\rangle$ and
$\vert\uparrow\downarrow\rangle$. Varying ${\bf\Omega}$ with
$\alpha_e=\alpha_d=1$ would lead to an over-counting of these states. 

Note that Eq.~(\ref{clos}) bears some similarities with the resolution
of the identity in the Hilbert space with no double occupancy
obtained from the spin-hole coherent states. \cite{Auerbach,Auerbach91}

\subsubsection{Trace}

The resolution of the identity (\ref{clos}) allows to compute the
trace of a given operator $\hat O$ in terms of the spin-particle-hole
coherent states. The derivation is standard\cite{Negele} and one obtains
\begin{equation}
{\rm Tr}\hat O = {\cal N} \int \frac{d{\bf \Omega}}{4\pi} \int
d\zeta^*d\zeta e^{-\alpha |\zeta|^2}\langle {\bf \Omega},\tilde\zeta
\vert\hat P\hat O \vert {\bf \Omega},\zeta\rangle , 
\label{trace}
\end{equation}
where $\tilde \zeta \equiv (-f_\uparrow,-f_\downarrow,e,p,d)$. 

As a simple application of Eq.~(\ref{trace}), we calculate the
partition function for a single site. Imposing the constraints
(\ref{cons}) by using Eq.~(\ref{coh1}), we easily obtain
\begin{eqnarray}
Z_{\rm at} &=& {\rm Tr} e^{-\beta (\hat H - \mu \hat N)} \nonumber \\
  &=& {\cal N} \Biggl( \frac{1}{\alpha_e^2\alpha_d} + \frac{e^{\beta
\mu}}{\alpha_e\alpha_d} + \frac{e^{\beta (2\mu-U)}}{\alpha_e\alpha_d^2}
  \Biggr) \nonumber \\ 
  &=& 1+2e^{\beta\mu}+e^{\beta(2\mu-U)} ,
\end{eqnarray}
where $\hat N$ is the total number of particles and $\beta=1/T$ is the
inverse temperature.

\subsection{Path integral for a single site}
\label{sec:piat}

In order to express the partition function as a path integral, we use
Eq.~(\ref{trace}) and  divide the ``time'' interval $\beta$ into $M$ steps:
\begin{eqnarray}
Z_{\rm at} &=& {\cal N} \int \frac{d{\bf \Omega}}{4\pi} \int
d\zeta^*d\zeta  e^{-\alpha |\zeta|^2} \nonumber \\ && \times
\langle {\bf \Omega},\tilde\zeta \vert \hat P e^{-\epsilon (\hat
H-\mu\hat N)}\cdots e^{-\epsilon (\hat H-\mu\hat N)} \vert {\bf
\Omega},\zeta\rangle , 
\label{Z0}
\end{eqnarray}
where $\epsilon=\beta/M$. We then introduce $(M-1)$ times the resolution of
the identity (\ref{clos}). The details of this procedure are given in Appendix
\ref{appI}. One finds 
\begin{eqnarray}
Z_{\rm at} &=& {\cal N}^M \int \Biggl( \prod_{k=1}^M
\frac{d{\bf\Omega}_k}{4\pi} 
d\zeta^*_kd\zeta_k d\lambda_k\Biggr) \nonumber \\ && \times
\exp \Biggl\lbrace \sum_{k=1}^M \Bigl [ -\alpha |\zeta_k|^2 
+\zeta^*_k\zeta_{k-1} \nonumber \\ && + f^*_{\uparrow k}p^*_k f_{\uparrow
k-1} p_{k-1}(\langle {\bf\Omega}_k | {\bf\Omega}_{k-1}\rangle -1)
-\epsilon K_{k,k-1} \Bigr 
] \Biggl \rbrace , \nonumber \\ && 
\label{Z1}
\end{eqnarray}
where 
\begin{eqnarray}
K_{k,k-1} &=& \frac{\langle {\bf\Omega}_k,\zeta_k\vert\hat K_k \vert
{\bf \Omega}_{k-1},\zeta_{k-1}\rangle}
{\langle{\bf\Omega}_k,\zeta_k\vert{\bf\Omega}_{k-1},\zeta_{k-1}\rangle} ,
\label{Kkm} \\ 
\hat K_k &=& \hat H-\mu\hat N +i \lambda^{(1)}_k \hat Q^{(1)}_k+i \sum_\sigma
\lambda^{(2)}_{\sigma k} \hat Q^{(2)}_\sigma .
\end{eqnarray}
$\lambda_k\equiv(\lambda^{(1)}_k,\lambda^{(2)}_{\sigma k})$ denotes
Lagrange multipliers which impose 
the constraints (\ref{cons}). 

An important difference with the standard path integral for a system of 
fermions and bosons comes from the presence of the constants
$\alpha_e=\alpha_d$  
($\neq 1$) which ensure that the empty and doubly occupied states
are not over-counted. Taking the continuum time limit ($\epsilon\to 0$) from
Eq.~(\ref{Z1}) would then lead to an infinite chemical potential
$\mu_e=-(\alpha_e-1)/\epsilon$ for the $e$ boson:
\begin{equation}
\sum_{k=1}^M e^*_k(e_{k-1}-\alpha_e e_k)
\to -\int_0^\beta d\tau e^*\Biggl( \partial_\tau  + \frac{\alpha_e-1}{\epsilon}
\Biggr) e , 
\end{equation}
and a similar result for the $d$ boson (i.e. $\mu_d=-(\alpha_d-1)/\epsilon$). 

At half-filling, $\mu=U/2$ due to particle-hole symmetry. The
contribution of the empty and doubly occupied states to the partition
function is exponentially small at low temperature ($T\ll U/2$) and
can be neglected. It is then possible to take $\alpha_e=\alpha_d=1$,
and the continuum time limit is well defined. This
argument  fails away from half-filling in the single-site case. However, we are
interested in the full Hubbard model, where the chemical potential is
determined by the filling of the Hubbard bands in the strong
correlation limit. In presence of
a small concentration of holes, $\mu\simeq W/2$ is located near the top
of the LHB (of total width $W$). In the low-temperature
limit $T\ll W$, the system in the atomic limit is in its ground-state
with exactly one particle per site. The contribution of the empty and
doubly occupied atomic states to the partition function is again negligible,
which allows to take $\alpha_e=\alpha_d=1$. This argument is particularly
clear in the strong-coupling expansion (Sec.~\ref{sec:sce}) where the
atomic action is used only in the calculation of the connected atomic
functions $G^{Rc}$. The latter are obtained from the atomic limit
($t=0$) of the full action, the chemical potential being determined
by the full Hubbard model.  

Taking $\alpha_e=\alpha_d=1$ and ignoring the overall normalization
constant ${\cal N}^M$,\cite{note1} we write the partition function as 
\begin{eqnarray}
Z_{\rm at} &=& \int \Biggl( \prod_{k=1}^M \frac{d{\bf\Omega}_k}{4\pi}
d\zeta^*_kd\zeta_k d\lambda_k\Biggr) \exp \Biggl\lbrace -\sum_{k=1}^M \Bigl
[ \zeta^*_k(\zeta_k-\zeta_{k-1}) \nonumber \\ && -
 f^*_{\uparrow k}p^*_k f_{\uparrow k-1} p_{k-1}(\langle {\bf\Omega}_k
| {\bf\Omega}_{k-1}\rangle -1) +\epsilon K_{k,k-1} \Bigr ] \Biggl
\rbrace . \nonumber \\ &&  
\label{Z2}
\end{eqnarray}
The continuum time limit can now be taken without difficulty
(see Appendix \ref{appI}):
\begin{equation}
Z_{\rm at} = \int {\cal D}{\bf\Omega} \int {\cal D}\lambda \int {\cal
D}[f,e,p,d] e^{-S_{\rm at}}, \nonumber 
\end{equation}
where the action is given by
\begin{eqnarray}
S_{\rm at} &=& S^{(0)}_{\rm at} + \int d\tau f^*_\uparrow p^* f_\uparrow p
\langle{\bf\Omega} | \dot{\bf\Omega} \rangle , \nonumber \\
S^{(0)}_{\rm at} &=& \int d\tau \Bigl[ -i\lambda^{(1)}+ \sum_\sigma
f^*_\sigma(\partial_\tau-\mu+i\lambda^{(2)}_\sigma)f_\sigma \nonumber \\ && 
+ e^*(\partial_\tau+i\lambda^{(1)})e +
p^*(\partial_\tau+i\lambda^{(1)}-i\lambda^{(2)}_\uparrow)p \nonumber \\ && 
+ d^*(\partial_\tau+U+i\lambda^{(1)}-i\sum_\sigma\lambda^{(2)}_\sigma)d \Bigr ] .
\label{action0}
\end{eqnarray}
$S^{(0)}_{\rm at}$ is similar to the action obtained from the
Kotliar-Ruckenstein slave bosons\cite{Kotliar86} with
$p_\uparrow\equiv p$ and without the $p_\downarrow$ boson. 
The difference between $S_{\rm at}$ and $S^{(0)}_{\rm at}$, which
arises from the dynamics of ${\bf\Omega}$, is nothing but the Berry
phase term\cite{Berry84} 
$A^0=\langle {\bf\Omega}|\dot{\bf\Omega}\rangle$ of a
spin-$\frac{1}{2}$ ($|\dot{\bf\Omega}\rangle=\partial_\tau
|{\bf\Omega}\rangle$). The Berry phase 
term is alive whenever the state is singly occupied ($f^*_\uparrow
f_\uparrow\equiv 1$ and $p^*p\equiv 1$).

\subsection{Path integral for the Hubbard model}
\label{sec:pihm}

The definition of the spin-particle-hole coherent states can be
extended to the lattice case:
\begin{equation}
\vert {\bf \Omega},\zeta\rangle = \exp \Bigl( \sum_{\bf r} (e_{\bf r}
\hat e^\dagger_{\bf r} + p_{\bf r} \hat p^\dagger_{\bf r} + 
d_{\bf r} \hat d^\dagger_{\bf r} - \sum_\sigma f_{{\bf r}\sigma}\hat
f^\dagger_{{\bf r}\sigma}) \Bigr) \vert {\rm vac}\rangle,
\end{equation}
where the unit vector ${\bf\Omega}_{\bf r}$ is now site dependent
[${\bf\Omega}=\lbrace {\bf\Omega}_{\bf r} \rbrace$ and
$\zeta=\lbrace\zeta_{\bf r} \rbrace$]. The fermionic
operator $\hat f_{{\bf r}\sigma}$ is defined with 
respect to the local spin-quantization axis ${\bf\Omega}_{\bf r}$. 

The procedure to write the partition function as a path integral is
similar to the single-site case. The only contribution which is not
purely atomic comes from the intersite hopping term. The latter
modifies $K_{k,k-1}$ (or, in the continuum time limit, $K_{k,k}$)
[Eq.~(\ref{Kkm})]. As shown in Appendix \ref{appII}
\begin{equation}
\langle {\bf\Omega}_k,\zeta_k\vert \hat c^\dagger_{\bf r} \hat c_{{\bf
r}'} \vert {\bf \Omega}_k,\zeta_k\rangle  = \gamma^\dagger_{{\bf r} k}
R^\dagger_{{\bf r} k} R_{{\bf r}'k}\gamma_{{\bf r}'k}
\langle{\bf\Omega}_k,\zeta_k\vert{\bf\Omega}_k,\zeta_k\rangle ,
\label{hop1}
\end{equation}
where $\hat c_{\bf r}=(\hat c_{{\bf r}\uparrow},\hat c_{{\bf
r}\downarrow})^T$.  
\begin{equation}
R_{{\bf r} k} = e^{-\frac{i}{2}\varphi_{{\bf r} k} \sigma_z}
         e^{-\frac{i}{2}\theta_{{\bf r} k} \sigma_y}
         e^{-\frac{i}{2}\psi_{{\bf r} k} \sigma_z}
\label{Rdef}
\end{equation}
is a SU(2)/U(1) matrix which rotates the spin-quantization axis from $\hat
{\bf z}$ to ${\bf\Omega}(\theta_{{\bf r} k},\varphi_{{\bf r}
k})$. It satisfies $R_{{\bf r} k} \sigma_z R_{{\bf r} k}^\dagger =
{\bf\Omega}_{{\bf r} k} \cdot \bbox{\sigma}$, where
$\bbox{\sigma}=(\sigma_x,\sigma_y,\sigma_z)$ stands for the Pauli matrices. 
$\gamma_{{\bf r} k}=(\gamma_{{\bf r}\uparrow k},\gamma_{{\bf
r}\downarrow k})^T$, and  
\begin{eqnarray}
\gamma_{{\bf r}\uparrow k} &=& e^*_{{\bf r} k} p_{{\bf r} k} f_{{\bf
r}\uparrow k} , \nonumber \\  
\gamma_{{\bf r}\downarrow k} &=& p^*_{{\bf r} k} d_{{\bf r} k} f_{{\bf
r}\downarrow k} . 
\label{gamdef1}
\end{eqnarray} 
The action of the Hubbard model then reads
\begin{equation}
S = S_{\rm at} - \sum_{{\bf r},{\bf r}'}\int d\tau \gamma_{\bf
r}^\dagger R^\dagger_{\bf r} 
t_{{\bf r}{\bf r}'} R_{{\bf r}'}\gamma_{{\bf r}'}, 
\label{action1}
\end{equation}
where $S_{\rm at}$ is the generalization to $N$ sites of the atomic
action given by Eqs.~(\ref{action0}). In the Hubbard model, $t_{{\bf
r}{\bf r}'}$ equals $t$ if 
${\bf r}$ and ${\bf r}'$ are nearest neighbors and vanishes otherwise. 
Since the constraints (\ref{cons}) are
preserved under the time evolution determined by the action $S$
[Eq.~(\ref{action1})], we can replace the functional integral over
$\lambda(\tau)$  by an integral over a set
$\lambda\equiv(\lambda^{(1)}_{\bf r},\lambda^{(2)}_{{\bf r}\sigma})$ of
time-independent Lagrange multipliers:
\begin{equation} 
\int {\cal D} \lambda \to \int d\lambda \equiv \prod_{\bf r} \Biggl
( \int_0^{\frac{2\pi}{\beta}} \frac{\beta d\lambda^{(1)}_{\bf r}}{2\pi}
\prod_\sigma \int_0^{\frac{2\pi}{\beta}} \frac{\beta d\lambda^{(2)}_{{\bf
r}\sigma}}{2\pi} 
\Biggr).
\end{equation}  

Equations (\ref{action0}) and (\ref{action1}) provide a
spin-rotation-invariant slave-boson formulation of the Hubbard
model. \cite{Li89} They have been derived recently without refering 
explicitly to the spin-particle-hole coherent states, but by introducing a
fluctuating spin-quantization axis in the functional integral. \cite{ND}

\section{Strong-coupling expansion}
\label{sec:sce}

In this section, we recast the spin-particle-hole coherent-state path
integral in a form suitable for a perturbative expansion
with respect to $t/U$. This is achieved by
performing two successive transformations of the intersite hopping
term (Secs.~\ref{sec:hs} and \ref{sec:ad}). The resulting action is
expressed as a function of two fermionic fields ($\gamma_\uparrow$ and
$\gamma_\downarrow$) which describe the propagation of particles in
the LHB and UHB, and a unit vector field ${\bf\Omega}$. In
Sec.~\ref{sec:ea}, we show that this action can be truncated to
quartic order in the fermionic fields in the strong correlation
limit. Finally, in Sec.~\ref{sec:dpt} we show how this effective
strong-coupling action can be used as the starting point of a
diagrammatic perturbative expansion with respect to $t/U$, and discuss
possible applications of our formalism.

\subsection{Grassmannian Hubbard-Stratonovich transformations}
\label{sec:hs}

Following
Refs.~\cite{Sarker88,Pairault98,ND} we decouple the intersite
hopping term in Eq.~(\ref{action1}) by means of a Grassmannian
Hubbard-Stratonovich transformation. The partition function becomes 
\begin{eqnarray}
Z &=& \int {\cal D}{\bf\Omega} {\rm det} (\hat t) \int {\cal D}[\psi] 
\exp \Bigl\lbrace -\sum_{a,b} \psi^*_a \hat t^{-1}_{ab} \psi_b
\Bigr\rbrace \nonumber \\ && \times
\int d\lambda \int {\cal D}[f,e,p,d] \exp \Bigl\lbrace -S_{\rm at} 
+\sum_a(\psi^*_a \gamma_a + {\rm c.c.}) \Bigr\rbrace , \nonumber \\ && 
\end{eqnarray}
where $\psi$ is an auxiliary fermionic field which couples to the
(composite) field $\gamma$ defined in Eq.~(\ref{gamdef1}). We use the notation
\begin{equation}
\psi_a \equiv \psi_{{\bf r}_a\sigma_a}(\tau_a), \,\,\,
\sum_a \equiv \sum_{{\bf r}_a,\sigma_a} \int d\tau_a ,
\end{equation}
and denote by
\begin{equation}
\hat t_{{\bf r}{\bf r}'} = R^\dagger_{\bf r} t_{{\bf r}{\bf r}'} R_{{\bf r}'}
\end{equation}
the ${\bf\Omega}$-dependent intersite hopping matrix. The SU(2)/U(1)
matrix $R_{\bf r}$ rotates the spin-quantization axis from $\hat{\bf z}$ to
${\bf\Omega}_{\bf r}$ [see Eq.~(\ref{Rdef})]. The diagonal elements of
$\hat t_{{\bf r}{\bf r}'}$  
correspond to intraband propagation, while its off-diagonal elements
describe transitions between the LHB and UHB. Note that ${\rm det} (\hat
t)$ is a function of ${\bf\Omega}$ and should therefore be kept explicitly in
the functional integral. 

Performing the functional integral over the fields $f,e,p,d$ and the
Lagrange multipliers $\lambda$, we obtain
\begin{eqnarray}
\int d\lambda \int {\cal D}[f,e,p,d] \exp \Bigl\lbrace -S_{\rm at} + 
\sum_a(\psi^*_a \gamma_a + {\rm c.c.}) \Bigr\rbrace && \nonumber \\ 
= Z_{\rm at}[{\bf\Omega}] e^{W[\psi^*,\psi;{\bf\Omega}]} && 
\label{W0}
\end{eqnarray}
where
\begin{eqnarray}
Z_{\rm at}[{\bf\Omega}] &=& \int d\lambda \int {\cal D}[f,e,p,d]
e^{-S_{\rm at}}
\nonumber \\ 
&=& Z^{(0)}_{\rm at} e^{-S_B[{\bf\Omega}]}   
\end{eqnarray}
is the ``partition function'' in the atomic limit ($t=0$) for a given
configuration of ${\bf\Omega}$. $Z^{(0)}_{\rm at}$ is the partition
function obtained from $S^{(0)}_{\rm at}$. 
We emphasize here that the integration of
the Lagrange multipliers and the bosonic fields has been done
exactly. $S_B[{\bf\Omega}]$ is the action of
the spin degrees of freedom in the atomic limit. Treating 
$A^0_{\bf r}=\langle{\bf\Omega}_{\bf r}|\dot{\bf\Omega}_{\bf r}\rangle$ in
perturbation, we obtain to lowest order (in a cumulant expansion)\cite{ND}
\begin{eqnarray}
S_B[{\bf\Omega}] &=& \sum_{\bf r} \int d\tau A^0_{\bf r} \langle
f^*_{{\bf r}\uparrow}p^*_{\bf r} 
f_{{\bf r}\uparrow}p_{\bf r} \rangle _{S^{(0)}_{\rm at}} \nonumber \\ 
&=& \sum_{\bf r} \int d\tau A^0_{\bf r} .
\end{eqnarray}
$S_B$ is a collection of Berry phase terms for spins localized at the
lattice sites. In Eq.~(\ref{W0}), $W[\psi^*,\psi;{\bf\Omega}]$ is the
generating functional of the connected atomic Green's functions 
\begin{eqnarray}
G^{Rc}_{\lbrace a_i,b_i\rbrace} &=&  (-1)^R \langle
\gamma_{a_1}\cdots\gamma_{a_R}\gamma_{b_R}^* \cdots \gamma_{b_1}^*
\rangle _{\rm at,c} \nonumber \\  
&=& \frac{\delta^{(2R)} W[\psi^*,\psi;{\bf\Omega}]}{\delta\psi^*_{a_1}\cdots
\delta \psi^*_{a_R}\delta \psi_{b_R}\cdots \delta \psi_{b_1}} 
\Biggr |_{\psi^*=\psi=0} .
\label{GRat}
\end{eqnarray}
Note that the $G^{Rc}$'s are defined for a given configuration of the
field ${\bf\Omega}$. 

The partition function is thus written as
\begin{eqnarray}
&& Z = Z^{(0)}_{\rm at} \int {\cal D}{\bf\Omega} {\rm det}(\hat t)
\int {\cal D}[\psi]  
e^{-S[\psi^*,\psi;{\bf\Omega}]} , \nonumber \\ 
&& S[\psi^*,\psi;{\bf\Omega}]  = S_B[{\bf\Omega}]+\sum_{a,b} \psi^*_a
\hat t^{-1}_{ab} \psi_b - W[\psi^*,\psi;{\bf\Omega}] .
\label{Spsi}
\end{eqnarray} 
$W[\psi^*,\psi;{\bf\Omega}]$
can be obtained explicitly by inverting Eq.~(\ref{GRat}):
\begin{equation}
W[\psi^*,\psi] = \sum_{R=1}^\infty \frac{(-1)^R}{(R!)^2}
\sum_{a_i,b_i}' \psi^*_{a_1}\cdots \psi^*_{a_R} \psi_{b_R}\cdots
\psi_{b_1} G^{Rc}_{\lbrace a_i,b_i\rbrace} .
\label{W2}
\end{equation}
The primed summation in (\ref{W2}) reminds us that all the fields in a
given product $\psi^*_{a_1}\cdots \psi_{b_1}$ share the same value of
the site index. 

In order to completely determine the action $S[\psi^*,\psi;{\bf\Omega}]$, we
need to compute the atomic Green's functions $G^{Rc}$ from the
action $S_{\rm at}$. The procedure
to obtain these quantities can be found in Appendix \ref{appIII} and in
Ref.~\onlinecite{ND}. Here we only quote the final results. The
single-particle Green's function is
\begin{equation}
G^{-1}_{{\bf r}\sigma} = G^{(0)-1}_\sigma - {\rm sgn}(\sigma) A^0_{\bf r} ,
\end{equation}
where ${\rm sgn}(\sigma)=1(-1)$ for
$\sigma=\uparrow(\downarrow)$. $G^{(0)}$ is the 
atomic Green's function corresponding to $S^{(0)}_{\rm at}$. In the low
temperature limit
\begin{eqnarray}
G^{(0)}_\uparrow(\tau) &=& \theta(-\tau+\eta) e^{\mu\tau} , \nonumber \\ 
G^{(0)}_\downarrow(\tau) &=& -\theta(\tau-\eta) e^{(\mu-U)\tau} ,
\label{G1}
\end{eqnarray}
where $\eta\to 0^+$. In Fourier space,
Eqs.~(\ref{G1}) become
$G^{(0)}_\uparrow(i\omega)=(i\omega+\mu)^{-1}$ and 
$G^{(0)}_\downarrow(i\omega)=(i\omega+\mu-U)^{-1}$ where $\omega$ is a
fermionic Matsubara frequency. As shown in
Sec.~(\ref{sec:ea}), the $R$-particle Green's functions $G^{Rc}$ ($R\geq 2$)
are involved only in (virtual) interband transitions. The typical
energy scale ($\sim U$) for the latter is much larger than the typical
energy scale for spin fluctuations ($\sim J=4t^2/U$) so that the
effect of the Berry phase term $A^0$ can be neglected (i.e. we can use
$S^{(0)}_{\rm at}$ instead of $S_{\rm at}$ to compute $G^{Rc}$ for
$R\geq 2$). We find 
\begin{equation}
G^{Rc}_{\sigma\cdots\sigma,\sigma\cdots\sigma}=0 \,\,\, (R\geq 2),
\end{equation}
which implies that there is no intraband interaction for the $\psi$
field. The two-particle atomic Green's function $G^{{\rm II}c}$ is thus
entirely determined by $G^{{\rm
II}c}_{\uparrow\downarrow,\uparrow\downarrow}$, or  equivalently by
the two-particle vertex (see Eq.~(\ref{GaII})) 
\begin{equation}
\Gamma^{\rm
II}_{\uparrow\downarrow,\uparrow\downarrow}(i\omega_1,i\omega_2;
i\omega_3(i\omega_4)) =i\omega_2-i\omega_3-U ,
\label{GaIIbis}
\end{equation}
where $\omega_4=\omega_1+\omega_2-\omega_3$ is fixed by energy
conservation. 
We shall show in Sec.~\ref{sec:ea} that the effective action in the
strong-coupling regime ($t/U\ll 1$)  is entirely determined by the
knowledge of $G$ and $G^{{\rm II}c}$ (or $\Gamma^{\rm II}$) so that we
do not require the knowledge of $G^{Rc}$ for $R\geq 3$.  

\begin{figure}
\epsfysize 5 cm
\epsffile[20 400 290 590]{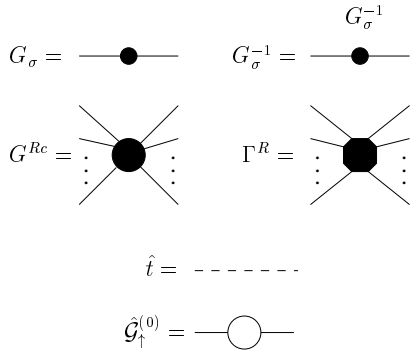}
\epsfysize 8 cm
\caption{ 
Definition of the various symbols appearing in the Feynman
diagrams. Connected atomic Green's functions are represented by a
filled circle and atomic vertices by polygons, with the exception of
$G^{-1}$ which is also shown as a filled circle. The intersite hopping
matrix $\hat t$ is represented by a dashed line, and the propagator
$\hat {\cal G}^{(0)}_\uparrow$ defined in Eq.~(\ref{Gcal0}) by an empty
circle. } 
\label{fig:def}
\end{figure}

\begin{figure}
\epsfysize 2 cm
\epsffile[0 355 260 427]{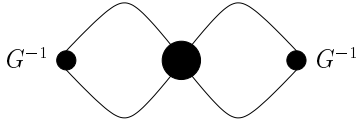}
\caption{A Feynman diagram contributing to $S'[{\bf\Omega}]$.  }
\label{fig:Sp}
\end{figure}

We could now follow Ref.~\onlinecite{Pairault98} and use the action
$S[\psi^*,\psi;{\bf\Omega}]$ to perform a diagrammatic perturbative
expansion with respect to the intersite hopping amplitude $t$. The
matrix structure of $\hat t_{{\bf r}{\bf r}'}$ distinguishes processes that do
conserve the number of doubly occupied sites from those that involve
interband transitions, thus allowing for a $t/U$ expansion. [In
Ref.~\onlinecite{Pairault98}, the auxiliary field 
$\psi_\sigma$ couples to the original fermionic
field $c_\sigma$. The resulting perturbative expansion involves two
dimensionless parameters, $t/U$ and $t/T$, and breaks down at low
temperature.] However, the drawback of using the action
$S[\psi^*,\psi;{\bf\Omega}]$ is that the field $\psi$ has no direct physical
meaning. It is rather the fields to which $\psi_\uparrow$ and
$\psi_\downarrow$ couple, namely the composite fields $e^*pf_\uparrow$ and
$p^*df_\downarrow$, which have a direct physical interpretation in terms of
particles propagating in the LHB and UHB. This suggests to perform a
second Hubbard-Stratonovich transformation of the intersite hopping
term. Denoting by $\gamma$ the auxiliary field of this transformation,
we rewrite the partition function as
\begin{eqnarray}
Z &=& Z^{(0)}_{\rm at}\int {\cal D}{\bf\Omega} \int {\cal D}[\gamma,\psi]
\exp\Bigl\lbrace-S_B[{\bf\Omega}] +\sum_{a,b}\gamma^*_a \hat t_{ab} \gamma_b
\nonumber \\ && 
+\sum_a (\gamma^*_a\psi_a + {\rm c.c.}) +W[\psi^*,\psi;{\bf\Omega}] 
\Bigr\rbrace  .
\end{eqnarray}
Integrating out the $\psi$ field, we obtain
\begin{eqnarray}
Z &=& Z^{(0)}_{\rm at} \int {\cal D}{\bf\Omega} \tilde Z[{\bf\Omega}] 
e^{-S_B[{\bf\Omega}]} , \nonumber \\ && \times
\int {\cal D}[\gamma] \exp \Bigl\lbrace \sum_{a,b} \gamma^*_a \hat t_{ab}
\gamma_b + \tilde W[\gamma^*,\gamma;{\bf\Omega}] \Bigr\rbrace  ,
\end{eqnarray}
Note that $\gamma$ is now an {\it elementary} field. [We expect
$\gamma$ to have the same physical meaning as the composite fields
defined in Eq.~(\ref{gamdef1}), hence the same notation.]  
$\tilde W[\gamma^*,\gamma;{\bf\Omega}]$ is the generating
functional of the 
connected Green's functions $\tilde G^{Rc}$ calculated with the action
$-W[\psi^*,\psi;{\bf\Omega}]$. Since the latter is local (in space),
so are the $\tilde G^{Rc}$'s. $\tilde Z[{\bf\Omega}]$ is defined by 
\begin{eqnarray}
\tilde Z[{\bf\Omega}] &=& \int {\cal D}[\psi] e^{W[\psi^*,\psi;{\bf\Omega}]}
\nonumber \\ &=& {\rm det}(-G) e^{-S'[{\bf\Omega}]} .
\end{eqnarray}
${\rm det}(-G)$ comes from the Gaussian part, $\sum_{a,b} G_{ab}\psi^*_a
\psi_b$, of the action $-W[\psi^*,\psi;{\bf\Omega}]$. Non-Gaussian terms give
the contribution $S'[{\bf\Omega}]$:
\begin{equation}
e^{-S'[{\bf\Omega}]} = \Biggl\langle \exp\Biggl\lbrace
\sum_{R=2}^\infty \frac{(-1)^R}{(R!)^2} \sum_{a_i,b_i}' \psi^*_{a_1}\cdots
\psi_{b_1} G^{Rc}_{\lbrace a_i,b_i\rbrace} \Biggr\rbrace
\Biggr\rangle ,
\end{equation}
where the average is to be taken with the action $\sum_{a,b} G_{ab}
\psi^*_a\psi_b$. $S'[{\bf\Omega}]$ can be represented as a sum of Feynman
diagrams, the bare propagator being $-G^{-1}$ and the vertices the
atomic Green's functions $G^{Rc}$ (with $R\geq 2$). An example of
diagram is given in Fig.~\ref{fig:Sp}. [The symbols used in the
Feynman diagrams are defined in Fig.~\ref{fig:def}.] This diagram, as
well as the other diagrams contributing to $S'[{\bf\Omega}]$ does not have
any physical meaning. As will be shown below, $S'[{\bf\Omega}]$ does
not contribute to the final result. ``Physical'' and ``anomalous''
(i.e. without ``physical'' meaning) diagrams can be defined rigorously
by considering the perturbation theory based on the action
$S[\psi^*,\psi;{\bf\Omega}]$ [Eq.~(\ref{Spsi})]. Within the latter,
all diagrams can be represented by atomic Green's functions connected
by the intersite hopping matrix $\hat t$. The perturbation theory
based on the action $S[\gamma^*,\gamma;{\bf\Omega}]$ generates
diagrams that cannot be represented in this way. These ``anomalous''
diagrams should not contribute to physical quantities. Indeed, we will
find that the total contribution of anomalous diagrams always vanishes
(see Sec.~\ref{sec:ad}).   

Noting that ${\rm det}(-G)=Z^{-1}_{\rm at}[{\bf\Omega}]$,\cite{note3} we obtain
\begin{eqnarray}
Z &=& \int {\cal D}{\bf\Omega} \int {\cal D}[\gamma] 
e^{-S[\gamma^*,\gamma;{\bf\Omega}]} , \nonumber \\
S[\gamma^*,\gamma;{\bf\Omega}] &=& S'[{\bf\Omega}]-\sum_{a,b}
\gamma^*_a \hat t_{ab} \gamma_b - \tilde W[\gamma^*,\gamma;{\bf\Omega}] .
\end{eqnarray}
In order to completely determine the action $S[\gamma^*,\gamma;{\bf\Omega}]$,
we need to compute the (local) Green's functions $\tilde G^{Rc}$ from
the action $-W[\psi^*,\psi;{\bf\Omega}]$ [Eq.~(\ref{W2})]. 

\begin{figure}
\epsfysize 2.8 cm
\epsffile[0 355 250 445]{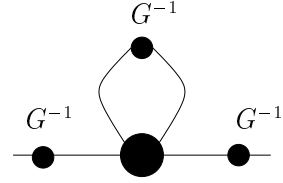}
\caption{A Feynman diagram contributing to $\tilde\Gamma$.}
\label{fig:gamt1}
\end{figure}

Let us first
consider the single-particle Green's function $\tilde G_{ab}=-\langle
\psi_a\psi^*_b\rangle_{-W}$. Retaining only the Gaussian part of
the action $-W$, i.e. $\sum_{a,b} G_{ab} \psi^*_a\psi_b$, one obtains
$\tilde G=-G^{-1}$. More generally, we have
\begin{equation}
\tilde G_{ab} = -G^{-1}_{ab} + \tilde\Gamma_{ab} ,
\label{Gt1}
\end{equation}
where $\tilde\Gamma$ can be represented as a sum of Feynman diagrams
arising from the 
vertices $G^{Rc}$ (with $R\geq 2$). An example is shown in
Fig.~\ref{fig:gamt1}. It is easy to see that all the diagrams
contributing to $\tilde\Gamma$ are anomalous. The fact that $\tilde
\Gamma$ has no physical meaning can be understood as follows. We can
view $\tilde\Gamma$ as a self-energy correction to $G$ [see
Eq.~(\ref{Gt1})]. Since $G$ is the exact atomic 
Green's function, we do not expect any local (i.e. atomic)
correction. We will see in
Sec.~\ref{sec:ad} that the role of $\tilde\Gamma$ is to cancel
other anomalous contributions arising in the strong-coupling
perturbative expansion. 

\begin{figure}
\epsfysize 7 cm
\epsffile[20 275 260 515]{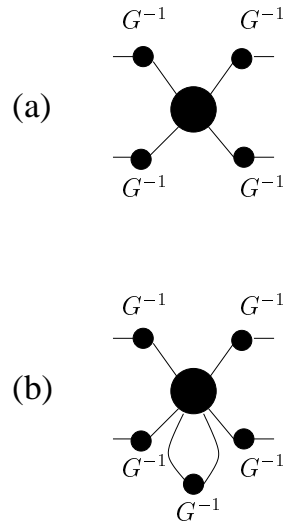}
\caption{Feynman diagrams for $\tilde G^{{\rm II}c}$. (a)
$-\Gamma^{\rm II}$. (b) Anomalous diagram.}
\label{fig:GIIt}
\end{figure}

\begin{figure}
\epsfysize 6 cm
\epsffile[160 250 450 540]{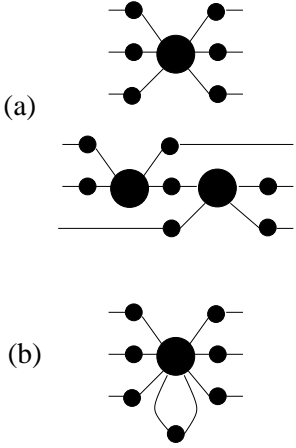}
\caption{ Examples of Feynman diagrams contributing to $\tilde
G^{{\rm III}c}$. (a) Contributions to $\Gamma^{\rm III}$. (b)
A contribution to the anomalous part $\tilde\Gamma^{\rm III}$.
All the filled circles with two external legs denote $G^{-1}$. }
\label{fig:GIIIt}
\end{figure}

A result similar to Eq.~(\ref{Gt1}) holds for higher-order Green's
functions. The simplest diagram for the two-particle Green's function $\tilde
G^{{\rm II}c}_{a_1a_2,b_1b_2}=\langle \psi_{a_1}\psi_{a_2}
\psi^*_{b_2}\psi^*_{b_1}\rangle_{-W}$
(shown in Fig.~\ref{fig:GIIt}a) gives the contribution $-\Gamma^{\rm
II}_{a_1a_2,b_1b_2}$, where $\Gamma^{\rm II}_{a_1a_2,b_1b_2}$ is the
two-particle vertex. All other diagrams are anomalous
(Fig.~\ref{fig:GIIt}b). In 
the same way, $\tilde G^{\rm IIIc}$ can be written as the sum of
$\Gamma^{\rm III}$ and an anomalous contribution $\tilde\Gamma^{\rm
III}$ (Fig.~\ref{fig:GIIIt}). More generally, we find
\begin{equation}
\tilde G^{Rc}_{\lbrace a_i,b_i\rbrace} = -(-1)^R \Bigl
( \Gamma^R_{\lbrace a_i,b_i\rbrace} +\tilde
\Gamma^{R}_{\lbrace a_i,b_i\rbrace} \Bigr) \,\,\,(R\geq 2), 
\end{equation}
where $\Gamma^R$ is the atomic $R$-particle vertex and $\tilde
\Gamma^{R}$ denotes anomalous contributions. 

We can then rewrite the action as
\begin{eqnarray}
&& S[\gamma^*,\gamma;{\bf\Omega}] = S'[{\bf\Omega}] - \sum_{a,b}
\gamma^*_a (\hat 
t_{ab}+G^{-1}_{ab}-\tilde\Gamma_{ab}) \gamma_b \nonumber \\ && 
+ \sum_{R=2}^\infty \frac {1}{(R!)^2} \sum_{a_i,b_i}' \Bigl
( \Gamma^R_{\lbrace a_i,b_i\rbrace} 
+ \tilde \Gamma^{R}_{\lbrace a_i,b_i\rbrace} \Bigr)
\gamma^*_{a_1} \cdots \gamma_{b_1} . \nonumber \\ && 
\label{action2}
\end{eqnarray} 
Note that this result has been obtained without any approximation and
gives the exact value of the partition function if one carries out the
functional integral over the fields $\gamma_\sigma$ and ${\bf\Omega}$. By
performing two successive Hubbard-Stratonovich transformations, we
have summed the atomic contributions to the single-particle propagator
and the $R$-particle vertices ($R\geq 2$). As a result, the action
$S[\gamma^*,\gamma;{\bf\Omega}]$ is essentially parametrized by the intersite
hopping matrix $\hat t$ and the atomic vertices $G^{-1}$ and
$\Gamma^R$ ($R\geq 2$). [We show in Sec.~\ref{sec:ad} how to deal with
the anomalous vertices $\tilde\Gamma^R$.] Using
$S[\gamma^*,\gamma;{\bf\Omega}]$ instead of 
$S[f,e,p,d,\lambda;{\bf\Omega}]$ [Eq.~(\ref{action1})] or
$S[\psi^*,\psi;{\bf\Omega}]$ [Eq.~(\ref{Spsi})] presents two main
advantages. First we now deal with only three fields whose physical
meaning is clear. $\gamma_\uparrow$ and $\gamma_\downarrow$ describe particles
propagating in the LHB and UHB, respectively, while the dynamics of
${\bf\Omega}$ arises from spin fluctuations. [Note however that the
Green's function of the original fermions (those involved in the
definition of the Hubbard model) is not merely the propagator
$-\langle \gamma_a\gamma_b^*\rangle$ (see Sec.~\ref{sec:dpt}).] Second,
$S[\gamma^*,\gamma;{\bf\Omega}]$ turns out to be a convenient starting point
in the strong correlation limit $t/U\ll 1$. Indeed, for $U\to\infty$
(no interband transition), only the Gaussian part of the action needs
to be considered. The lowest-order corrections in $t/U$ are determined
by a small number of atomic vertices. In practice,
$S[\gamma^*,\gamma;{\bf\Omega}]$ can be truncated to quartic order in
the fermionic fields and is then
parametrized only by $G^{-1}$ and $\Gamma^{\rm II}$ (Sec.~\ref{sec:ea}).

\subsection{Anomalous diagrams}
\label{sec:ad}

In order to understand the role of anomalous diagrams, it is useful to
first consider the atomic limit ($t=0$) of the action
(\ref{action2}). In that limit, one has to recover the known (exact)
results. 

Let us consider first the partition function. Integrating out the
$\gamma$ field, one obtains
\begin{eqnarray}
Z_{\rm at}[{\bf\Omega}] &=& {\rm det}(-G^{-1})
e^{-S'[{\bf\Omega}]-S''[{\bf\Omega}]} ,
\nonumber \\   
e^{-S''[{\bf\Omega}]} &=& \Biggl\langle \exp \Biggl\lbrace -\sum_{a,b}
\gamma^*_a \tilde \Gamma_{ab} \gamma_b \nonumber \\ && 
- \sum_{R=2}^\infty 
\frac {1}{(R!)^2} \sum_{a_i,b_i}' \Bigl 
( \Gamma^R_{\lbrace a_i,b_i\rbrace} + \tilde \Gamma^{R}_{\lbrace
a_i,b_i\rbrace } \Bigr)
\gamma^*_{a_1} \cdots \gamma_{b_1} \Biggr\rbrace \Biggr\rangle 
\nonumber \\ &&
\label{Spp} 
\end{eqnarray}
where the average is to be taken with the Gaussian action
$-\sum_{a,b} \gamma^*_aG^{-1}_{ab} \gamma_b$. In order to obtain the
correct result
\begin{eqnarray}
Z_{\rm at} &=& \int {\cal D}{\bf\Omega} {\rm det}(-G^{-1}) \nonumber
\\ &=& Z_{\rm at}^{(0)} \int {\cal D}{\bf\Omega} e^{-S_B[{\bf\Omega}]} ,
\end{eqnarray}
one must have $S'[{\bf\Omega}]+S''[{\bf\Omega}]=0$. This result can be
obtained by inspection. Consider for instance the diagrams containing
once and only once the vertex $\Gamma^{\rm II}$ (and no higher-order
vertex). The diagram contributing to $S'[{\bf\Omega}]$ is shown in
Fig.~\ref{fig:Sp} and has the overall factor $-1/2$. There are two
contributions to $S''[{\bf\Omega}]$. They come from the terms
$\langle\gamma^*_a \tilde\Gamma_{ab}\gamma_b\rangle$ and $\langle
\Gamma^{\rm II}_{a_1a_2,b_1b_2}\gamma^*_{a_1}\gamma^*_{a_2}
\gamma_{b_2}\gamma_{b_1}\rangle$ where the average is taken with
the action $-\sum_{a,b}\gamma^*_aG^{-1}_{ab}\gamma_b$ [see
Eq.~(\ref{Spp})]. Both contributions correspond to the diagram of
Fig.~\ref{fig:Sp}, but with the factors $1$ and $-1/2$. We conclude
that the total contribution to $S'+S''$ vanishes.  

In the same way, we can calculate the single-particle Green's function
$-\langle \gamma_a\gamma^*_b\rangle$ (for a given configuration of
${\bf\Omega}$) and verify that we do obtain the
correct result in the atomic limit, namely $G_{ab}$. From the action
(\ref{action2}), we read off the inverse (atomic) propagator
$G^{-1}-\tilde\Gamma$. Self-energy corrections due to vertices $R\geq
2$ should therefore cancel $\tilde\Gamma$. Again this result can be
proved by inspection. Consider again the diagrams containing once the
vertex $\Gamma^{\rm II}$. The contribution to $\tilde\Gamma$ is shown
in Fig.~\ref{fig:gamt1}. The one-loop self-energy correction coming
from $\Gamma^{\rm II}$ (see Eq.~(\ref{action2})) produces the same
diagram but with opposite sign. 

\begin{figure}
\epsfysize 5.5 cm
\epsffile[0 270 260 495]{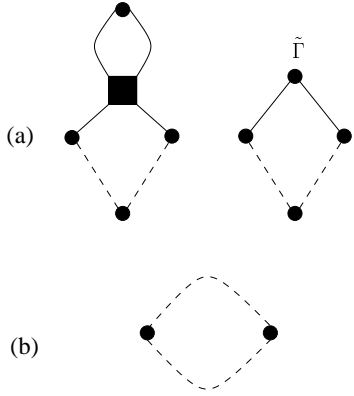}
\caption{(a) Two anomalous diagrams for the free energy.
Their sum vanishes when $\tilde\Gamma$ is approximated by the
contribution shown diagrammatically in Fig.~\ref{fig:gamt1}. 
(b) The corresponding ``physical'' diagram.}
\label{fig:Zan}
\end{figure}

Thus we come to the conclusion that anomalous diagrams cancel in the
atomic limit. We expect this property to hold also when $t\neq 0$, since
anomalous diagrams do not have any physical meaning. Consider for
instance the anomalous diagrams shown in Fig.~\ref{fig:Zan}a which
contribute to the effective action of the spin degrees of freedom
$S[{\bf\Omega}]=-\ln Z[{\bf\Omega}]$. 
[Fig.~\ref{fig:Zan}b shows the corresponding
``physical'' diagram.] Working out signs and symmetry factors, we find
that the sum of the two diagrams of Fig.~\ref{fig:Zan}a vanishes if
$\tilde\Gamma$ is approximated by the contribution shown
diagrammatically in Fig.~\ref{fig:gamt1}. 

As expected on physical grounds, anomalous diagrams always cancel in
the calculation of physical quantities. We can then work with the
action
\begin{eqnarray}
S[\gamma^*,\gamma;{\bf\Omega}] &=& - \sum_{a,b} \gamma^*_a (\hat
t_{ab}+G^{-1}_{ab}) \gamma_b \nonumber \\ && 
+ \sum_{R=2}^\infty \frac {1}{(R!)^2} \sum_{a_i,b_i}'
 \Gamma^R_{\lbrace a_i,b_i\rbrace} 
\gamma^*_{a_1} \cdots \gamma_{b_1} 
\label{action3}
\end{eqnarray} 
provided that we discard all anomalous diagrams.

\subsection{Effective action of the Hubbard model in the
strong correlation limit}
\label{sec:ea}

In this section, we show that only the quadratic and quartic parts of
the action $S[\gamma^*,\gamma;{\bf\Omega}]$ need to be considered in the
strong-coupling limit. This is achieved by considering the effective
action $S_{\rm LHB}[\gamma^*_\uparrow,\gamma_\uparrow;{\bf\Omega}]$ of
particles propagating in the LHB. To first order in $t/U$, $S_{\rm
LHB}[\gamma^*_\uparrow,\gamma_\uparrow;{\bf\Omega}]$ should correspond
to the $t$-$J$ model. This result will guide us in the determination of
the leading terms in the action of the Hubbard model in the strong correlation
limit. We emphasize however that our primary goal is not to determine
the effective action of carriers in the LHB, but to derive an
effective action for the full Hubbard model in the strong correlation
limit. As discussed at the end of this section, there are advantages
in working with the Hubbard model instead of considering the $t$-$J$
model obtained by integrating out the UHB.   

$S_{\rm LHB}$ is obtained by integrating out the $\gamma_\downarrow$
field:\cite{ND} 
\begin{equation}
e^{-S_{\rm LHB}[\gamma^*_\uparrow,\gamma_\uparrow;{\bf\Omega}]} = \int
{\cal D}[\gamma_\downarrow] e^{-S[\gamma^*,\gamma;{\bf\Omega}]} .
\label{Slhb}
\end{equation}

At zeroth order in $t/U$, interband transitions are neglected. Since
$\Gamma^R_{\sigma\cdots\sigma}=0$, diagrams contributing to $S^{(0)}_{\rm LHB}$
must necessary contain closed loops of $\gamma_\downarrow$ particle. In the
absence of particles in the UHB, these loops vanish (see Appendix
\ref{appIV}). As a result, the
effective action of the LHB is Gaussian to leading order in $t/U$:
\begin{eqnarray}
S^{(0)}_{\rm LHB} &=& \sum_{\bf r} \int d\tau \gamma^*_{{\bf
r}\uparrow} (\partial_\tau -\mu  
+A^0_{\bf r}) \gamma_{{\bf r}\uparrow} \nonumber \\ && 
- \sum_{{\bf r},{\bf r}'}\int d\tau \gamma^*_{{\bf r}\uparrow} \hat
t_{{\bf r}\uparrow,{\bf r}'\uparrow} \gamma_{{\bf r}'\uparrow} . 
\label{Slhb0}
\end{eqnarray}
The action (\ref{Slhb0}) describes particles propagating in the LHB and
interacting with spin fluctuations {\it via} the gauge field $A^0$ and the
${\bf\Omega}$-dependent intersite hopping matrix $\hat t$
(i.e. $S^{(0)}_{\rm LHB}$ describes fermions coupled to a U(1) gauge
field). A detailed discussion of its physical meaning can be found in
Sec.~III.B.1 of Ref.~\onlinecite{ND}. 

\begin{figure}
\epsfysize 8 cm
\epsffile[60 350 300 640]{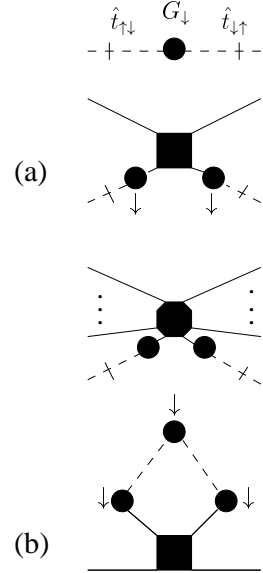}
\caption{(a) Contributions of order $t/U$ to the effective action
$S_{\rm LHB}$. (b) An example of vanishing diagram. Slashed dashed
lines indicate interband transitions. }
\label{fig:Slhb}
\end{figure}

We now consider the contributions of order $t/U$ to $S_{\rm LHB}$
[Eq.~(\ref{Slhb})]. They are represented by the diagrams of
Fig.~\ref{fig:Slhb}a. Each of these diagrams contains two interband
transitions. The vanishing of closed loops with $\gamma_\downarrow$
particles ensures that all the diagrams to be considered are of the
type shown in Fig.~\ref{fig:Slhb}a. Examples of vanishing diagrams are shown in
Fig.~\ref{fig:Slhb}b.

Let us first focus on the quadratic and quartic contributions
generated by the integration of the UHB (first two diagrams of
Fig.~\ref{fig:Slhb}a). These terms are non-local in time.
However, since they involve (virtual) interband transitions, they can be
approximated by local vertices. From the equations of motion in the
atomic limit, we deduce $\gamma_\uparrow(\tau')=
e^{-\mu(\tau-\tau')}\gamma_\uparrow(\tau)$.\cite{Zinn}  Using the latter
equation and integrating over the time differences, the non-local vertices are
then approximated by local vertices. For instance, the two-point
vertex (first diagram of Fig.~\ref{fig:Slhb}a) is approximated as 
\begin{eqnarray}
&& \sum_{{\bf r},{\bf r}',{\bf r}''} \int d\tau d\tau' \gamma^*_{{\bf
r}\uparrow}(\tau) \hat 
t_{{\bf r}\uparrow,{\bf r}'\downarrow}(\tau) G^{(0)}_\downarrow(\tau-\tau') \hat
t_{{\bf r}'\downarrow,{\bf r}''\uparrow}(\tau') \gamma_{{\bf
r}''\uparrow}(\tau') 
\nonumber \\ && \simeq 
-\frac{1}{U} \sum_{{\bf r},{\bf r}',{\bf r}''} \int d\tau
\gamma^*_{{\bf r}\uparrow}(\tau) \hat 
t_{{\bf r}\uparrow,{\bf r}'\downarrow}(\tau) t_{{\bf r}'\downarrow,{\bf
r}''\uparrow}(\tau) \gamma_{{\bf r}''\uparrow}(\tau),
\end{eqnarray}
where we have used Eq.~(\ref{G1}). Since we can ignore the effect of
spin fluctuations on (virtual) interband transitions, we have replaced
$G_\downarrow$ by $G^{(0)}_\downarrow$. If we also approximate the
quartic term by a 
local vertex and neglect higher-order vertices, we end up with an
effective action which is nothing but the action of the $t$-$J$ model
(including the so-called pair-hopping terms) in the spin-hole
coherent-state path integral.\cite{note3.1} [See
Ref.~\onlinecite{ND} for a detailed derivation.] It is crucial
here that the effective action is local in time,  otherwise it would not
derive from an Hamiltonian. It should be noted that a similar
approximation is made in the standard derivation of the $t$-$J$ model,
which starts from the resolvant operator $\hat R(E)=(E-\hat H)^{-1}$. 
Projecting out states with double occupancy yields an effective
``Hamiltonian'' $\hat H_{\rm eff}(E)$ which depends on the energy $E$. The
Hamiltonian of the $t$-$J$ model is obtained by retaining terms of
order $O(t/U)$ and replacing $E$ by the energy $E_0$ in the absence of
interband coupling. \cite{Auerbach,Fulde} In our formalism, this last
step amounts to approximating the vertices of $S_{\rm LHB}$ by local
vertices (in time) using the atomic equations of motion for the
$\gamma_\uparrow$ field.  

Since the quadratic and quartic terms of the action $S_{\rm LHB}$ are
sufficient to recover the action of the $t$-$J$ model, we expect
higher-order vertices to vanish at the same level of
approximation. Consider 
the contribution of order $t/U$ to $S_{\rm LHB}$ due to $\Gamma^{\rm
III}_{\uparrow\uparrow\downarrow,\uparrow\uparrow\downarrow}$
(Fig.~\ref{fig:Slhb}a). This term involves the product  
\begin{equation}
\gamma^*_{{\bf r}\uparrow}(\tau_1)\gamma^*_{{\bf r}\uparrow}(\tau_2)\gamma^*_{{\bf r}'\uparrow}(\tau_3)
\gamma_{{\bf r}''\uparrow}(\tau_3')\gamma_{{\bf r}\uparrow}(\tau_2')\gamma_{{\bf r}\uparrow}(\tau_1'),
\end{equation}
where the fields are evaluated at different times and three different
sites (${\bf r}$, ${\bf r}'$ and ${\bf r}''$). If we now
approximate this term by using the atomic equation of motion, we
obtain the product $[\gamma^*_{{\bf r}\uparrow}(\tau)]^2\gamma^*_{{\bf
r}'\uparrow}(\tau) 
\gamma_{{\bf r}''\uparrow}(\tau)[\gamma_{{\bf r}\uparrow}(\tau)]^2$
which vanishes since it involves squares of Grassmann variables.
The same reasoning holds for higher-order vertices.  
Therefore, using the same approximations that lead to the $t$-$J$  
model, only the quadratic and quartic terms of the action $S_{\rm LHB}$
subsist.

We conclude that in the strong-coupling limit, we can truncate the
action (\ref{action3}) to quartic order in the fermionic fields:
\bleq
\begin{eqnarray}   
S[\gamma^*,\gamma;{\bf\Omega}] &=& \sum_{\bf r} \int d\tau
\gamma^*_{{\bf r}\uparrow}  
(\partial_\tau-\mu+A^0_{\bf r}) \gamma_{{\bf r}\uparrow} 
-\sum_{{\bf r},{\bf r}'} \int d\tau \gamma^*_{{\bf r}\uparrow} 
\hat t_{{\bf r}\uparrow,{\bf r}'\uparrow} \gamma_{{\bf r}'\uparrow} 
\nonumber \\ && 
+\sum_{\bf r} \int d\tau \gamma^*_{{\bf r}\downarrow} 
(\partial_\tau-\mu+U-A^0_{\bf r}) \gamma_{{\bf r}\downarrow} 
-\sum_{{\bf r},{\bf r}'} \int d\tau \gamma^*_{{\bf r}\downarrow} 
\hat t_{{\bf r}\downarrow,{\bf r}'\downarrow} \gamma_{{\bf r}'\downarrow} 
\nonumber \\ && 
-\sum_{{\bf r},{\bf r}'} \int d\tau \Bigl( \gamma^*_{{\bf r}\uparrow} 
\hat t_{{\bf r}\uparrow,{\bf r}'\downarrow} \gamma_{{\bf
r}'\downarrow} +{\rm c.c.} \Bigr)  
\nonumber  \\ && 
+\sum_{\bf r} \int d\tau_1 d\tau_2  d\tau_3 d\tau_4
\Gamma^{\rm
II}_{\uparrow\downarrow,\uparrow\downarrow}(\tau_1,\tau_2;\tau_3,\tau_4)  
\gamma^*_{{\bf r}\uparrow}(\tau_1)\gamma^*_{{\bf r}\downarrow}(\tau_2) 
\gamma_{{\bf r}\downarrow}(\tau_4)\gamma_{{\bf r} \uparrow}(\tau_3) .
\label{action4}
\end{eqnarray} 
\eleq
Eq.~(\ref{action4}) is the main result of this paper. It gives the
effective action of the Hubbard model in the strong-coupling limit $t/U\ll
1$. The fields $\gamma_\uparrow$ and $\gamma_\downarrow$ correspond to
particles propagating in the LHB and the UHB. They are coupled by the
interband hopping matrix $\hat t_{\uparrow\downarrow}$ and the  
two-particle atomic vertex 
$\Gamma^{\rm II}_{\uparrow\downarrow,\uparrow\downarrow}$, and
interact with the spin fluctuations which show up in the dynamics of
${\bf\Omega}$. Note that one can describe the interaction with spin
fluctuations by a SU(2) gauge field with $A^0$ its time component.

The action (\ref{action4}) is suitable for a
perturbative calculation of the free energy and the LHB Green's
functions. However, except at half-filling, it does not allow to
compute the UHB Green's functions. In a hole doped system, closed
loops of $\gamma_\uparrow$ particles do not vanish (see Appendix
\ref{appIV}) so that the arguments used above for the LHB do not hold
for the UHB. [Even at zeroth order in $t/U$, the effective action
$S_{\rm UHB}$ contains an infinite number of terms.] 

An interesting aspect of this action is that the strong local Coulomb
repulsion is already present in the quadratic part of the action. To
understand the physical meaning of $\Gamma^{\rm
II}_{\uparrow\downarrow,\uparrow\downarrow}$, we rewrite
Eq.~(\ref{GaIIbis}) for real frequencies ($i\omega\to\omega$) as
\begin{equation}
E_{\rm UHB}-E_{\rm LHB}=U+\Gamma^{\rm
II}_{\uparrow\downarrow,\uparrow\downarrow} , 
\label{dE}
\end{equation}
where we have identified $\omega_2$ ($\omega_3$) with the energy of
the particle in the UHB (LHB). Eq.~(\ref{dE}) suggests that we can
interpret $\Gamma^{\rm II}_{\uparrow\downarrow,\uparrow\downarrow}$ as
the residual 
interaction (given that a 
part of the interaction is included in the quadratic action) between
two particles siting at the same site. Alternatively, one can view the
$U$ term in the action as a mean field, and $\Gamma^{\rm
II}_{\uparrow\downarrow,\uparrow\downarrow}$ as the correction to this
mean field.  

Finally, we note that the action (\ref{action4}) without the quartic
term is similar to that obtained within the large-$U$ Hartree-Fock
approximation where the SU(2) spin-rotation invariance is maintained
by introducing a fluctuating spin-quantization axis
${\bf\Omega}$ in the functional integral.\cite{Schulz90} Omitting the
quartic term ($\Gamma^{\rm 
II}_{\uparrow\downarrow,\uparrow\downarrow}$) is however in general
not possible without missing processes of order $t/U$. For instance,
this term plays a central role in the derivation of the $t$-$J$
model.\cite{ND}

\subsection{Strong-coupling diagrammatic perturbation theory}
\label{sec:dpt}

The action (\ref{action4}) can be used as the starting point for a
perturbative calculation with respect to $t/U$. We discuss in the
following the computation of the free energy 
and the LHB Green's function to first order in $t/U$. Formally, we have 
\begin{eqnarray}
Z &=& \int {\cal D}{\bf\Omega} Z[{\bf\Omega}] , \nonumber \\ 
Z[{\bf\Omega}] &=& e^{-S[{\bf\Omega}]} = \int {\cal D}[\gamma]
e^{-S[\gamma^*,\gamma;{\bf\Omega}]} , 
\end{eqnarray}
and
\begin{eqnarray}
{\cal G}^{\rm LHB}_\sigma(a,b) &=& \frac{1}{Z} \int {\cal D}{\bf\Omega}
e^{-S[{\bf\Omega}]} (R_a)_{\sigma\uparrow} \hat {\cal G}_\uparrow(a,b)
(R^\dagger_b)_{\uparrow\sigma} , \label{Gcal1}  \\ 
\hat {\cal G}_\uparrow(a,b) &=& -\frac{1}{Z[{\bf\Omega}]} \int {\cal D}[\gamma]
\gamma_{a\uparrow} \gamma^*_{b\uparrow} e^{-S[\gamma^*,\gamma;{\bf\Omega}]} ,
\label{Gcal2}
\end{eqnarray} 
where $S[{\bf\Omega}]$ is the action for the spin degrees of
freedom and $\hat{\cal G}_\uparrow(a,b)$ the Green's function for a
given configuration of ${\bf\Omega}$. Note that the Green's function
of the original fermions $c$ (those involved in the definition of the
Hubbard model) is not merely $-\langle
\gamma_a\gamma^*_b\rangle$ (see Appendix
\ref{AppV}). Eq.~(\ref{Gcal1}) shows that the fermionic 
field $c^{\rm LHB}$ in the LHB can be expressed as
\begin{eqnarray}
c^{\rm LHB}_{{\bf r}\sigma} &=& (R_{\bf r})_{\sigma\uparrow}
\gamma_{{\bf r}\uparrow}  \\ 
&=& z_{{\bf r}\sigma} \gamma_{{\bf r}\uparrow} ,
\label{Clhb}
\end{eqnarray}
where the last line is obtained by writing the SU(2)/U(1) rotation
matrix $R_{\bf r}$ as
\begin{equation}
R_{\bf r} = 
\left( \begin{array}{cccc}
z_{{\bf r}\uparrow} & -z^*_{{\bf r}\downarrow} \\ 
z_{{\bf r}\downarrow}  & z^*_{{\bf r}\uparrow} 
\end{array} \right ),
\label{Schwinger}
\end{equation}
with the constraint $|z_{{\bf r}\uparrow}^2|+|z_{{\bf
r}\downarrow}^2|=1$. Eq.~(\ref{Clhb}) is familiar from the
Schwinger-boson slave-fermion formulation of the $t$-$J$
model. \cite{note5.0} 

\begin{figure}
\epsfysize 3.5 cm
\epsffile[10 323 260 465]{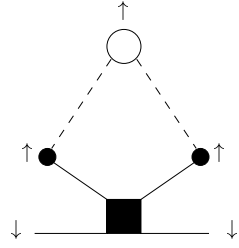}
\caption{ An example of self-energy diagram for the UHB particles
which is $O(1)$ in $t/U$. } 
\label{fig:SEuhb}
\end{figure}

We now introduce the propagator
\begin{equation}
\hat {\cal G}^{(0)-1}_\sigma = G^{-1}_\sigma + \hat t_{\sigma\sigma} .
\label{Gcal0}
\end{equation}
To zeroth order in $t/U$, the LHB Green's function is $\hat {\cal
G}^{(0)}_\uparrow$, and the partition function is given by
\begin{equation}
Z^{(0)}[{\bf\Omega}] = {\rm det} \Bigl(-\hat{\cal G}^{(0)-1}_\uparrow\Bigr) .
\label{ZOm0}
\end{equation}
As discussed in Sec.~\ref{sec:ea} $\hat{\cal G}^{(0)}_\downarrow$ is
not the UHB Green's function to 
zeroth order in $t/U$ except at half-filling. Even in the absence of
interband transition, 
there are $O(1)$ corrections to $\hat{\cal G}^{(0)}_\downarrow$
due to the non-vanishing of closed loops of $\gamma_\uparrow$ particles in
the (hole) doped system (see Appendix \ref{appIV}). A finite $O(1)$ 
self-energy correction to $\hat{\cal G}^{(0)}_\downarrow$ is 
shown in Fig.~\ref{fig:SEuhb}.  

\begin{figure}
\epsfysize 5 cm
\epsffile[150 255 450 535]{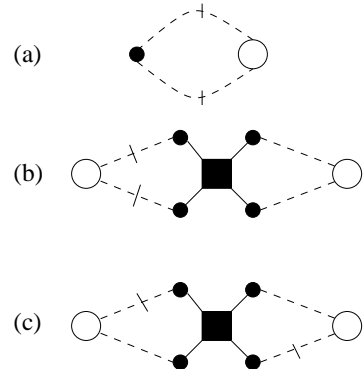}
\caption{
Diagrammatic representation of the effective action of the spin
degrees of freedom at order $O(t/U)$. (a) $S_1[{\bf\Omega}]$. (b)
$S_1'[{\bf\Omega}]$. (c) $S_1''[{\bf\Omega}]$. Slashed dashed lines indicate
interband transitions. } 
\label{fig:FE}
\end{figure}

To first order in $t/U$, there are three contributions to $S[{\bf\Omega}]$
shown diagrammatically in Fig.~\ref{fig:FE}:
\bleq 
\begin{eqnarray} 
S_1[{\bf\Omega}] &=&  \sum_{a,b,a',b'} \hat{\cal G}^{(0)}_\uparrow
(a,b) \hat t_{\uparrow\downarrow}(b,b') G_\downarrow
(b',a') \hat t_{\downarrow\uparrow}(a,a') , \nonumber \\
S_1'[{\bf\Omega}] &=& \sum_{a,b,a',b'} \sum_{a_i,b_i}  
\Gamma^{\rm II}_{\uparrow\downarrow,\uparrow\downarrow}(a,b;a',b') 
G_\uparrow(a',a_4) \hat t_{\uparrow\uparrow}(a_4,a_3) 
\hat {\cal G}^{(0)}_\uparrow(a_3,a_2) \hat t_{\uparrow\uparrow}(a_2,a_1)
G_\uparrow(a_1,a) \nonumber \\ && \times
G_\downarrow(b',b_4) 
\hat t_{\downarrow\uparrow}(b_4,b_3)
\hat {\cal G}^{(0)}_\uparrow(b_3,b_2)\hat t_{\uparrow\downarrow}(b_2,b_1)
G_\downarrow(b_1,b) , \nonumber \\ 
S_1''[{\bf\Omega}] &=& - \sum_{a,b,a',b'} \sum_{a_i,b_i} 
\Gamma^{\rm II}_{\uparrow\downarrow,\uparrow\downarrow}(a,b;b',a') 
G_\uparrow(b',b_4) \hat t_{\uparrow\uparrow}(b_4,b_3) 
\hat {\cal G}^{(0)}_\uparrow(b_3,b_2) \hat t_{\uparrow\downarrow}(b_2,b_1)
G_\downarrow(b_1,b) \nonumber \\ && \times 
G_\downarrow(a',a_4) \hat t_{\downarrow\uparrow}(a_4,a_3)
\hat {\cal G}^{(0)}_\uparrow(a_3,a_2)\hat t_{\uparrow\uparrow}(a_2,a_1)
G_\uparrow(a_1,a) .
\end{eqnarray}  
\eleq

The effective action $S[{\bf\Omega}]$ of the spin degrees of freedom
to order $O(t/U)$ can be
computed exactly at half-filling. The zeroth-order contribution
[Eq.~(\ref{ZOm0})] yields the Berry phase term $S_B$ to lowest order in
$A^0$. When computing $S_1$, $S_1'$ and $S_1''$ we may replace $\hat {\cal
G}^{(0)}_\uparrow$ by $G_\uparrow$, since the neglected terms will be of
higher-order in $t/U$.\cite{note5} Furthermore, since $S_1$, $S_1'$
and $S_1''$ describe
(virtual) interband transitions, we can ignore the effect of the Berry
phase term and replace $G_\uparrow$ by $G^{(0)}_\uparrow$. We then find 
\begin{eqnarray}
S_1[{\bf\Omega}] &=& - \frac{1}{U} \sum_{{\bf r},{\bf r}'} \int d\tau \hat
t_{{\bf r}\uparrow,{\bf r}'\downarrow} \hat t_{{\bf r}'\downarrow,{\bf
r}\uparrow} \nonumber \\ 
&=& \frac{1}{2U} \sum_{{\bf r},{\bf r}'} t^2_{{\bf r}{\bf r}'} \int
d\tau ({\bf\Omega}_{\bf r} 
\cdot {\bf\Omega}_{{\bf r}'}-1) .
\label{Som1}
\end{eqnarray}
In Eq.~(\ref{Som1}), the exchange interaction is assumed to be
instantaneous, since the characteristic spin fluctuation energy
$J=4t^2/U$ is much smaller than $U$. The last line of Eq.~(\ref{Som1})
is obtained from $\hat t_{{\bf r}\uparrow,{\bf
r}'\downarrow} \hat 
t_{{\bf r}'\downarrow,{\bf r}\uparrow}=(1-{\bf\Omega}_{\bf r} \cdot
{\bf\Omega}_{{\bf r}'})/2$.\cite{ND} Since
$S_1'$ vanishes at half-filling, while $S_1''$ turns out to be of
order $O(t^3/U^3)$, 
we verify that the effective action of the spin degrees of freedom at
order $O(t/U)$, $S_B+S_1$,  is nothing but the action of the (quantum)
AF Heisenberg model expressed in terms of spin coherent states:
\begin{equation}
S_{\rm Heis}[{\bf\Omega}] =
S_B[{\bf\Omega}]+J\sum_{\langle {\bf r},{\bf r}' \rangle}
\int d\tau\, \Bigl(\frac{{\bf\Omega}_{\bf r}\cdot
{\bf\Omega}_{{\bf r}'}}{4}-\frac{1}{4} \Bigr) .
\label{Heis}
\end{equation}

\begin{figure}
\epsfysize 7 cm
\epsffile[190 200 450 580]{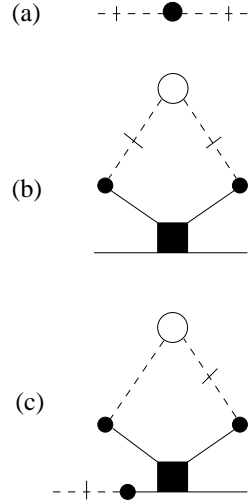}
\caption{
Diagrammatic representation of the LHB self-energy at order
$O(t/U)$. (a) $\hat\Sigma_1$. (b) $\hat\Sigma_1'$. (c) $\hat\Sigma_1''$ (the
symmetric diagram is not shown). Slashed dashed lines
indicate interband transitions. }
\label{fig:SE}
\end{figure}

The self-energy $\hat\Sigma$ of particles in the LHB is defined by
$\hat {\cal G}^{-1}_\uparrow =\hat {\cal G}^{(0)-1}_\uparrow
-\hat\Sigma$, where $\hat {\cal G}_\uparrow$ is the propagator for a
given configuration of ${\bf\Omega}$. It has three contributions
to first order in $t/U$ (Fig.~\ref{fig:SE}):
\bleq
\begin{eqnarray}
\hat\Sigma_1(a,a') &=& \sum_{b,b'} \hat
t_{\uparrow\downarrow}(a,b) G_\downarrow(b,b') \hat
t_{\downarrow\uparrow}(b',a') , \nonumber \\
\hat\Sigma_1'(a,a') &=& \sum_{b,b',b_i} \Gamma^{\rm
II}_{\uparrow\downarrow,\uparrow\downarrow}(a,b;a',b')
G_\downarrow(b',b_1) \hat t_{\downarrow\uparrow}(b_1,b_2) 
\hat {\cal G}^{(0)}_\uparrow(b_2,b_3)
\hat t_{\uparrow\downarrow}(b_3,b_4) G_\downarrow(b_4,b) , 
\nonumber \\ 
\hat\Sigma_1''(a,a') &=& \sum_{b,b'} \sum_{a_i,b_i} 
\hat t_{\uparrow\downarrow}(a,a_1) G_\downarrow(a_1,a_2)
\Gamma^{\rm II}_{\uparrow\downarrow,\uparrow\downarrow}(b,a_2;a',b')
G_\downarrow(b',b_1) \hat t_{\downarrow\uparrow}(b_1,b_2) 
\hat {\cal G}^{(0)}_\uparrow(b_2,b_3) \hat t_{\uparrow\uparrow}(b_3,b_4) 
G_\uparrow(b_4,b) \nonumber \\ && + \sum_{b,b'} \sum_{a_i,b_i} 
\Gamma^{\rm II}_{\uparrow\downarrow,\uparrow\downarrow}(a,b;b',a_2)
G_\uparrow(b',b_1) \hat t_{\uparrow\uparrow}(b_1,b_2) 
\hat {\cal G}^{(0)}_\uparrow(b_2,b_3) \hat t_{\uparrow\downarrow}(b_3,b_4) 
G_\downarrow(b_4,b)G_\downarrow(a_2,a_1) \hat t_{\downarrow\uparrow}(a_1,a') .
\end{eqnarray}
\eleq
$\hat\Sigma_1$ describes indirect hopping processes between
next-nearest-neighbor sites which occur {\it via} virtual transitions
to the UHB. In the $t$-$J$ model, these processes show up in the
pair-hopping term.  

So far our discussion has been confined to a formal framework. For
practical calculations, further approximations will be
necessary. Whereas the fermionic fields $\gamma_\sigma$ can be integrated
out in a systematic $t/U$ expansion, one still has to carry out the
functional integral over the unit vector field ${\bf\Omega}$. 

The simplest approach consists in expanding around a
broken-symmetry ground-state by making a saddle point approximation on
the spin variables ${\bf\Omega}_{\bf r}$. Different choices,
corresponding to AF, ferromagnetic or spiral orders,
are possible. This approach, which is
discussed at length in Ref.~\onlinecite{ND1}, can be justified by
taking a ``large-$S$'' semiclassical limit of the Hubbard model. The
spin-$\frac{1}{2}$ coherent states can be promoted to spin-$S$
coherent states (with $S$ arbitrary) by writing the SU(2)/U(1) rotation matrix
$R_{\bf r}$ as in Eq.~(\ref{Schwinger})
and generalizing the constraint $|z_{{\bf r}\uparrow}|^2+|z_{{\bf
r}\downarrow}|^2=1$ to $|z_{{\bf r}\uparrow}|^2+|z_{{\bf
r}\downarrow}|^2=2S$. The latter equality
allows to write $z_{{\bf r}\uparrow}=\sqrt{2S}\cos(\theta_{\bf r}/2)
e^{-\frac{i}{2}(\varphi_{\bf r}+\psi_{\bf r})}$ and $z_{{\bf
r}\downarrow}=\sqrt{2S}\sin(\theta_{\bf r}/2) 
e^{\frac{i}{2}(\varphi_{\bf r}-\psi_{\bf r})}$ where the choice of
$\psi_{\bf r}$ is free and corresponds to the U(1) gauge freedom. The
action of the Hubbard model is obtained from (\ref{action4}) with the
replacement $A^0\to 2SA^0$ and $\hat t\to 2S\hat t$. [Note that the
creation of a (fermionic) hole corresponds to a total removal of the
local moment.] In
the limit $S\to\infty$ (with $tS={\rm constant}$), the Berry phase
term suppresses quantum fluctuations of ${\bf\Omega}$. The spin
variables become classical and do not fluctuate at zero
temperature. The leading correction in $1/S$ gives the spin-wave modes
around the broken-symmetry ground-state.  

At half-filling, one can also try to compute $\cal G$ by directly
averaging $\hat {\cal G}$ with $S[{\bf\Omega}]$
[see Eq.~(\ref{Gcal1})], since the effective action of the spin
degrees of freedom, $S_{\rm Heis}[{\bf\Omega}]$, can be analyzed by various
methods. \cite{Auerbach} Since $S_{\rm Heis}[{\bf\Omega}]$ is not a
Gaussian action, this seems {\it a priori} a very difficult task. We
show in Ref.~\onlinecite{ND2} how this technical difficulty can be
circumvented by using the Schwinger-boson mean-field theory of the
Heisenberg model. The
advantage of this approach is that one starts from a 
good description of the magnetic properties of the system (which are
described by $S_{\rm Heis}$): absence of
long-range order at finite temperature, exponential divergence in
$1/T$ of the magnetic correlation length, etc. Moreover, this
formulation is a good starting point to study how spin fluctuations
affect hole motion.\cite{ND2} 

It should be pointed out that Eqs.~(\ref{Gcal1}) and (\ref{Gcal2})
cast the Hubbard model in the form of a spin-fermion model, which can
be seen as the strong-coupling counterpart of the ``weak-coupling''
spin-fermion model that has been studied so far. \cite{SF}
There are  however
important differences between the latter and our formalism: (i) In the
``weak-coupling'' spin-fermion model, the effective action of the spin
degrees of freedom is generally assumed to be Gaussian. In our
approach, the effective action $S[{\bf\Omega}]$ is obtained exactly
(to order $O(t/U)$) by integrating out the fermionic degrees of
freedom. (ii) The spin-fermion model omits coupling between spin
fluctuations and charge degrees of
freedom. Fermions interact with spin fluctuations {\it via} a spin-spin
interaction. In our approach, there is an intimate coupling between
charge degrees of freedom (i.e. hole motion) and spin
fluctuations, which shows up in the ${\bf\Omega}$-dependence of the
intersite hopping matrix. (iii) The spin-fermion model is restricted to the
weak-coupling regime, whereas our approach is appropriate to the
strong-coupling regime. \cite{note6}

Finally, we would like to discuss the relation to the $t$-$J$ model. 
Starting from the effective action (\ref{action4}) and integrating 
out the UHB, one recovers the action of the $t$-$J$ model in the
spin-hole coherent-state path integral. \cite{note3.1} A further
change of variables yields the action of the $t$-$J$ model in the
slave fermion formalism.\cite{Lee89} [See Sec.~III.B.3 of
Ref.~\onlinecite{ND}.] The connection between the action
(\ref{action4}) and the $t$-$J$ model is however not
straightforward and involves some subtleties regarding the proper time
ordering in the functional integral.\cite{ND} As a result, both
formulations should be considered as two different descriptions of the
strong-coupling 
limit of the Hubbard model. For instance, indirect hopping processes
between next-nearest neighbors appear in the Gaussian part of the action
(\ref{action4}), while they are described by the pair-hopping term in
the $t$-$J$ model (this term involves a three-body interaction in the
standard formulation of the $t$-$J$ model). But the most interesting
aspect of our formulation is the use of spin-particle-hole coherent
states. By clearly distinguishing between charge and spin degrees of
freedom, spin-particle-hole coherent states open up new possibilities
for analyzing hole motion in presence of strong spin
fluctuations. \cite{ND2}

\section{Summary and Conclusion}

There are two main ingredients in the derivation of the Heisenberg
model from the Hubbard model at half-filling: First, one should
deal with spin operators (instead of the original fermionic operators)
in order to account for the local moments that appear in the Mott
insulating phase. Then, the effective Hamiltonian for these spin
operators is derived by expanding with respect to
$t/U$. \cite{Anderson59} 

In this paper, we have proposed a generalization of this procedure to
the doped case. Our approach is based on the introduction of 
spin-particle-hole coherent states which generalize the
spin-$\frac{1}{2}$ coherent states by allowing the creation of a hole
or an additional particle. They also generalize
the spin-hole coherent states introduced in the context of the $t$-$J$
model.\cite{Auerbach91} The spin-particle-hole coherent states can be
used to derive a path integral and naturally lead to a
spin-rotation-invariant slave-boson formulation of the Hubbard model
(Sec.~\ref{sec:sph}). By performing two successive
Hubbard-Stratonovich transformations of the intersite hopping term,
the path integral can be recast in a from which is suitable for a
perturbation expansion in $t/U$. The action
$S[\gamma^*,\gamma;{\bf\Omega}]$ can be expressed in terms of two
Grassmann variables ($\gamma_\uparrow,\gamma_\downarrow$) and a unit
vector field. A singly occupied site is described by a
spin-$\frac{1}{2}$ coherent state $|{\bf\Omega}_{\bf r}\rangle$, while
$\gamma_\uparrow$ ($\gamma_\downarrow$) describes a particle
propagating in the LHB (UHB). The fermionic action has interaction
terms to all orders, given by the atomic vertices $\Gamma^R$
[Eq.~(\ref{action3})]. The fermions also interact with spin
fluctuations which show up in the dynamics of ${\bf\Omega}$. In the
strong-coupling limit $t\ll U$, the action can be truncated to quartic
order in the fermionic field [Eq.~(\ref{action4})] and used as the
starting point of a diagrammatic expansion. At half-filling and to order
$O(t/U)$, we recover the action of the (quantum) Heisenberg model. 

Since the fermionic field can be systematically integrated out within
a $t/U$ expansion, the main practical difficulty comes from the
dynamics of ${\bf\Omega}$ (spin fluctuations). The simplest approach
consists in expanding about a broken-symmetry ground-state my making a
saddle point approximation on the spin variables ${\bf\Omega}_{\bf
r}$. Different choices, corresponding to AF, ferromagnetic, or spiral
orders are possible. This approach is discussed in detail in
Ref.~\onlinecite{ND1}. The spin-particle-hole coherent-state path
integral turns out to be a very convenient tool for studying spin
waves about a broken-symmetry ground-state in the strong-coupling
limit. 

The main characteristic of the spin-particle-hole coherent-state path
integral lies in the clear distinction between charge ($\gamma$) and
spin (${\bf\Omega}$) degrees of freedom. This opens up new
possibilities for the computation of the Green's functions. One can
first calculate the Green's functions for a given (time-dependent)
configuration of ${\bf\Omega}$ and then perform the average with the
effective action $S[{\bf\Omega}]$ of the spin degrees of freedom [see
Sec.~\ref{sec:dpt} and Eqs.~(\ref{Gcal1}-\ref{Gcal2})]. This program
is carried out in Ref.~\onlinecite{ND2}.


\bleq 

\appendix

\section{Path integral for the single-site Hubbard model}
\label{appI}

Introducing $(M-1)$ times the resolution of the identity (\ref{clos}) in
Eq.~(\ref{Z0}), one obtains
\begin{equation}
Z_{\rm at} = {\cal N}^M \int \Biggl( \prod_{k=1}^M
\frac{d{\bf\Omega}_k}{4\pi} d\zeta^*_kd\zeta_k \Biggr)
e^{-\sum_{k=1}^M \alpha |\zeta_k|^2} \prod_{k=1}^M
\langle {\bf\Omega}_k,\zeta_k\vert \hat P_k e^{-\epsilon (\hat
H-\mu\hat N)} \vert {\bf \Omega}_{k-1},\zeta_{k-1}\rangle , 
\end{equation}
with $\epsilon=\beta/M$ and the boundary conditions:
${\bf\Omega}_0={\bf\Omega}_M$, $e_M=e_0$, $p_M=p_0$, $d_M=d_0$, and 
$f_{\sigma M}=-f_{\sigma 0}$. The projection operator $\hat P_k$ is defined by
Eq.~(\ref{proj}) with the spin-quantization axis determined by the
unit vector ${\bf\Omega}_k$. It can be written as
\begin{equation}
\hat P_k = \int_0^\frac{2\pi}{\epsilon} \frac{\epsilon d\lambda^{(1)}_k}{2\pi}
e^{-i\epsilon\lambda^{(1)}_k\hat Q^{(1)}_k}\prod_\sigma
\int_0^\frac{2\pi}{\epsilon} \frac{\epsilon d\lambda^{(2)}_{\sigma
k}}{2\pi} e^{-i\epsilon\lambda^{(2)}_{\sigma k}\hat Q^{(2)}_{\sigma k}} . 
\end{equation}
Since the constraints $\hat Q^{(1)}$ and $\hat Q^{(2)}_\sigma$ commute with the
Hamiltonian $\hat H-\mu\hat N$, we obtain
\begin{eqnarray} 
Z_{\rm at} &=& {\cal N}^M \int \Biggl( \prod_{k=1}^M
\frac{d{\bf\Omega}_k}{4\pi} d\zeta^*_kd\zeta_k d\lambda_k\Biggr) e^{-
\sum_{k=1}^M \alpha |\zeta_k|^2} \prod_{k=1}^M \langle
{\bf\Omega}_k,\zeta_k\vert \hat P_k e^{-\epsilon\hat K_k} \vert {\bf
\Omega}_{k-1},\zeta_{k-1}\rangle \label{A3} \\  
&\simeq& {\cal N}^M \int \Biggl( \prod_{k=1}^M \frac{d{\bf\Omega}_k}{4\pi}
d\zeta^*_kd\zeta_k d\lambda_k\Biggr) e^{-\sum_{k=1}^M \alpha
|\zeta_k|^2} \prod_{k=1}^M
\langle{\bf\Omega}_k,\zeta_k\vert{\bf\Omega}_{k-1},\zeta_{k-1}\rangle
e^{-\epsilon K_{k,k-1}} 
\end{eqnarray}
in the limit $\epsilon\to 0$, where
\begin{eqnarray}
\int d\lambda_k &\equiv& \int_0^\frac{2\pi}{\epsilon}\frac{\epsilon
d\lambda^{(1)}_k}{2\pi} \prod_\sigma\int_0^\frac{2\pi}{\epsilon} \frac{\epsilon
d\lambda^{(2)}_{\sigma k}}{2\pi} , \nonumber \\ 
\hat K_k &=& \hat H-\mu\hat N +i \lambda^{(1)}\hat Q^{(1)}_k+i \sum_\sigma
\lambda^{(2)}_{\sigma k} 
\hat Q^{(2)}_\sigma , \nonumber \\ K_{k,k-1} &=& \frac{\langle
{\bf\Omega}_k,\zeta_k\vert \hat K_k
\vert {\bf
\Omega}_{k-1},\zeta_{k-1}\rangle}{\langle{\bf\Omega}_k,
\zeta_k\vert{\bf\Omega}_{k-1},\zeta_{k-1}\rangle} .  
\end{eqnarray}
Using the expression of the scalar product [Eq.~(\ref{scalar2})], the
partition function is finally written as
\begin{equation}
Z_{\rm at} = {\cal N}^M \int \Biggl( \prod_{k=1}^M \frac{d{\bf\Omega}_k}{4\pi}
d\zeta^*_kd\zeta_k d\lambda_k\Biggr) \exp \Biggl\lbrace \sum_{k=1}^M \Bigl
[ -\alpha |\zeta_k|^2 +\zeta^*_k\zeta_{k-1} + f^*_{\uparrow k}p^*_k f_{\uparrow
k-1} p_{k-1}(\langle {\bf\Omega}_k | {\bf\Omega}_{k-1}\rangle -1)
-\epsilon  K_{k,k-1} \Bigr
] \Biggl \rbrace .
\label{ZZ1}
\end{equation}

Taking $\alpha_e=\alpha_d=1$ and neglecting the overall normalization
constant ${\cal N}^M$ (see the discussion in section \ref{sec:piat}), 
there is now no difficulty to take the continuum time
limit. $K_{k,k-1}\simeq K_{k,k}$ is evaluated by writing the Hamiltonian as
\begin{equation} 
\hat H-\mu\hat N = U\hat d^\dagger\hat d - \mu \sum_\sigma \hat
f^\dagger_\sigma\hat f_\sigma , 
\end{equation}
where the spin-quantization axis ${\bf\Omega}_k$ is chosen. The constraints
$\hat Q^{(1)}$ and $\hat Q^{(2)}_\sigma$ being also defined with respect to the
spin-quantization axis ${\bf\Omega}_k$, we obtain
\begin{eqnarray}
K_{k,k} &=& U |d_k|^2-\mu \sum_\sigma f^*_{\sigma k} f_{\sigma k}
+i\lambda^{(1)}_k(|e_k|^2+|p_k|^2+|d_k|^2-1) \nonumber \\ &&
+i\lambda^{(2)}_{\uparrow k}(f^*_{\uparrow k}f_{\uparrow
k}-|p_k|^2-|d_k|^2) 
+i\lambda^{(2)}_{\downarrow k}(f^*_{\downarrow k}f_{\downarrow k}-|d_k|^2) .
\end{eqnarray}
We also note that $\langle {\bf\Omega}_k|{\bf\Omega}_{k-1}\rangle-1=\langle
{\bf\Omega}_k|{\bf\Omega}_{k-1}\rangle-\langle
{\bf\Omega}_k|{\bf\Omega}_k\rangle$  yields a term 
$-\langle {\bf\Omega}|\dot{\bf\Omega}\rangle$ in the continuum time
limit. Here the 
dot denotes a time derivative. The final expression of the action 
$S_{\rm at}$ in the atomic limit is given by Eqs.~(\ref{action0}).

\section{Evaluation of \protect \small $\langle {\bf \Omega},\zeta \vert \hat
c^\dagger_{\bf r}\hat c_{{\bf r}'}\vert {\bf \Omega},\zeta\rangle$}
\label{appII}

In this appendix, we prove Eq.~(\ref{hop1}) (we do not write
explicitly the time index $k$). We introduce the operator
\begin{equation}
\hat\phi^\dagger_{{\bf r}\sigma} = \sum_{\sigma'}(R_{\bf
r})_{\sigma'\sigma}\hat c^\dagger_{{\bf r}\sigma'} 
\label{B1}
\end{equation}
which creates a particle with spin $\sigma$ in the spin
reference frame determined by ${\bf\Omega}_{\bf r}$. In the space
spanned by ${\cal B}_{\lbrace\bf\Omega_{\bf r}\rbrace}'$, we have
$\hat\phi_{{\bf r}\sigma} = \hat\gamma_{{\bf r}\sigma}$. We then deduce
\begin{eqnarray}
\langle {\bf \Omega},\zeta \vert \hat c^\dagger_{\bf r}\hat c_{{\bf
r}'}\vert {\bf \Omega},\zeta\rangle &=& \sum_{\sigma,\sigma'} \langle
{\bf \Omega},\zeta \vert 
\hat \gamma^\dagger_{{\bf r}\sigma}(R^\dagger_{\bf r} R_{{\bf
r}'})_{\sigma\sigma'} \hat 
\gamma_{{\bf r}'\sigma'} \vert {\bf \Omega},\zeta\rangle \nonumber \\ 
&=& \gamma^\dagger_{\bf r} R^\dagger_{\bf r} R_{{\bf r}'} \gamma_{{\bf
r}'} \langle 
{\bf\Omega},\zeta|{\bf\Omega},\zeta \rangle  ,
\label{appII.1}
\end{eqnarray}
where $\gamma_{{\bf r}\uparrow}=e^*_{\bf r} p_{\bf r} f_{{\bf r}\uparrow}$ and 
$\gamma_{{\bf r}\downarrow}=p^*_{\bf r} d_{\bf r} f_{{\bf r}\downarrow}$. The
intersite term in the action (\ref{action1}) follows from
Eq.~(\ref{appII.1}).

\section{Green's functions in the atomic limit}
\label{appIII}

In this section we compute the connected atomic Green's functions 
\begin{equation}
G^{Rc}_{\lbrace a_i,b_i\rbrace} =(-1)^R \langle T_\tau \hat\gamma_{a_1}
\cdots \hat\gamma_{a_R}\hat\gamma^\dagger_{b_R} \cdots \hat\gamma^\dagger_{b_1}
\rangle_{\rm at,c} , 
\end{equation}
written here in the operator formalism ($T_\tau$ is the imaginary-time
ordering operator). We consider a single site and drop the site index. 

Let us first consider the Green's functions determined by
$S^{(0)}_{\rm at}$. In the basis ${\cal B}_{\bf\Omega}'$, the
operators $\hat\gamma_\sigma$ and the Hamiltonian $\hat H-\mu\hat N$
can be written as (in matrix form)
\begin{eqnarray}
\hat\gamma_{\uparrow} &=& 
\left( \begin{array}{cccc}
0 & 1 & 0  \\ 0 & 0 & 0  \\ 0 & 0 & 0  
\end{array} \right ) ,
\,\,\,\, 
\hat\gamma_{\downarrow} = 
\left( \begin{array}{cccc}
0 & 0 & 0  \\ 0 &  0 & -1 \\ 0 & 0 & 0 
\end{array} \right ),  \nonumber \\ 
\hat H-\mu\hat N &=&  \left( \begin{array}{cccc}
0 & 0 & 0  \\ 0 & -\mu & 0  \\ 0 & 0 & U-2\mu   
\end{array} \right ) .
\end{eqnarray}
We easily find the time dependence of the operators
$\hat\gamma^{(\dagger)}_\sigma(\tau)=\hat
U(-\tau)\hat\gamma^{(\dagger)}_\sigma \hat U(\tau)$
(where $\hat U(\tau)=e^{-\tau(\hat H-\mu\hat N)}$ is the evolution operator):
\begin{eqnarray}
\hat\gamma_\uparrow(\tau) &=& e^{\mu\tau}\hat\gamma_\uparrow , \,\,\, 
\hat\gamma^\dagger_\uparrow(\tau) = e^{-\mu\tau}\hat\gamma^\dagger_\uparrow ,  
\nonumber \\ 
\hat\gamma_\downarrow(\tau) &=& e^{(\mu-U)\tau}\hat\gamma_\downarrow ,
\,\,\, \hat\gamma^\dagger_\downarrow(\tau) =
e^{-(\mu-U)\tau}\hat\gamma^\dagger_\downarrow .
\label{gamt}
\end{eqnarray}

The single-particle Green's function $G_\sigma^{(0)}(\tau)=-\langle T_\tau
\hat\gamma(\tau)\hat\gamma^\dagger_\sigma(\tau) \rangle$ is found to be
\begin{eqnarray}
G^{(0)}_\uparrow (\tau) &=& \frac{e^{\mu\tau}}{Z^{(0)}_{\rm at}}
[\theta(-\tau)e^{\beta\mu}-\theta(\tau)] \simeq
e^{\mu\tau}\theta(-\tau) , \nonumber \\  
G^{(0)}_\downarrow (\tau) &=& \frac{e^{(\mu-U)\tau}}{Z^{(0)}_{\rm at}}
[\theta(-\tau)e^{\beta(2\mu-U)}-\theta(\tau)e^{\beta\mu}] \simeq
-e^{(\mu-U)\tau}\theta(\tau)  .
\label{Gat1}
\end{eqnarray}
The final expressions are obtained in the low-temperature limit and we
have used $Z^{(0)}_{\rm at}=1+e^{\beta\mu}+e^{\beta(2\mu-U)}\simeq
e^{\beta\mu}$. The latter approximation neglects the contribution of
the empty and doubly occupied sites, in agreement with the discussion
of Sec.~\ref{sec:piat}. As expected, $Z^{(0)}_{\rm at}$ is the
partition function for a given spin direction. In Fourier space, we obtain
\begin{eqnarray}
G^{(0)}_\uparrow (i\omega) &=& (i\omega+\mu)^{-1} , \nonumber \\
G^{(0)}_\downarrow (i\omega) &=& (i\omega+\mu-U)^{-1} .
\end{eqnarray}

In the same way, we can calculate the two-particle Green's
function. For instance, we have
\begin{eqnarray}
G^{\rm II}_{\uparrow\uparrow,\uparrow\uparrow}(\tau_1,\tau_2;\tau_3,\tau_4) &=&
\frac{1}{Z^{(0)}_{\rm at}} e^{\mu(\tau_1+\tau_2-\tau_3-\tau_4)}{\rm
Tr}[\hat U(\beta) (\hat\gamma^\dagger_\uparrow \hat\gamma_\uparrow)^2] 
\nonumber \\ && \times 
[\theta(\tau_4-\tau_2)\theta(\tau_2-\tau_3)\theta(\tau_3-\tau_1) 
-\theta(\tau_3-\tau_2)\theta(\tau_2-\tau_4)\theta(\tau_4-\tau_1)
\nonumber \\ &&  
-\theta(\tau_4-\tau_1)\theta(\tau_1-\tau_3)\theta(\tau_3-\tau_2) 
+\theta(\tau_3-\tau_1)\theta(\tau_1-\tau_4)\theta(\tau_4-\tau_2)] ,
\end{eqnarray}
where ${\rm Tr}[\hat U(\beta) (\hat\gamma^\dagger_\uparrow
\hat\gamma_\uparrow)^2] =e^{\beta\mu}$. 
We do not distinguish between $G^{{\rm II}(0)}$ and
$G^{\rm II}$, since the effect of the Berry phase term on $G^{Rc}$
($\geq 2$) can
be ignored (see Sec.~\ref{sec:sce}). One can verify that $G^{{\rm
II}}_{\uparrow\uparrow,\uparrow\uparrow} (\tau_1,\tau_2;\tau_3,\tau_4)$
coincides with its disconnected part
\begin{eqnarray}
G^{{\rm II}dis}_{\uparrow\uparrow,\uparrow\uparrow}
(\tau_1,\tau_2;\tau_3,\tau_4) &=&  
G^{(0)}_\uparrow(\tau_1-\tau_3)G^{(0)}_\uparrow(\tau_2-\tau_4) - 
G^{(0)}_\uparrow(\tau_1-\tau_4)G^{(0)}_\uparrow(\tau_2-\tau_3) \nonumber \\
&=& e^{\mu(\tau_1+\tau_2-\tau_3-\tau_4)} 
[\theta(-\tau_1+\tau_3)\theta(-\tau_2+\tau_4)-\theta(-\tau_1+\tau_4)\theta(-\tau_2+\tau_3)] ,
\end{eqnarray}
so that $G^{{\rm II}c}_{\uparrow\uparrow,\uparrow\uparrow}=0$. 
Similarly $G^{{\rm
II}c}_{\downarrow\downarrow,\downarrow\downarrow}=0$. By using the
definition of the third cumulant
\begin{eqnarray}
G^{{\rm III}c}_{a_1a_2a_3,b_1b_2b_3} &=&
\frac{\delta^{(6)}W[\psi^*,\psi;{\bf\Omega}]}{\delta\psi^*_{a_1} \cdots
\delta\psi_{b_1}}\Biggr\vert_{\psi^*=\psi=0} \nonumber \\
&=& G^{\rm III}_{a_1a_2a_3,b_1b_2b_3}
- G^{(0)}_{a_1b_1} G^{{\rm II}c}_{a_2a_3,b_2b_3} 
- G^{(0)}_{a_2b_2} G^{{\rm II}c}_{a_1a_3,b_1b_3} 
- G^{(0)}_{a_3b_3} G^{{\rm II}c}_{a_1a_2,b_1b_2} 
+ G^{(0)}_{a_2b_1} G^{{\rm II}c}_{a_1a_3,b_2b_3}\nonumber \\ && 
- G^{(0)}_{a_3b_1} G^{{\rm II}c}_{a_1a_2,b_2b_3}
+ G^{(0)}_{a_3b_2} G^{{\rm II}c}_{a_1a_2,b_1b_3} 
+ G^{(0)}_{a_1b_2} G^{{\rm II}c}_{a_2a_3,b_1b_3} 
+ G^{(0)}_{a_2b_3} G^{{\rm II}c}_{a_1a_3,b_1b_2}
- G^{(0)}_{a_1b_3} G^{{\rm II}c}_{a_2a_3,b_1b_2} \nonumber \\ &&
- G^{(0)}_{a_1b_1} G^{(0)}_{a_2b_2}G^{(0)}_{a_3b_3}  
+ G^{(0)}_{a_1b_2} G^{(0)}_{a_2b_1}G^{(0)}_{a_3b_3} 
+ G^{(0)}_{a_1b_1} G^{(0)}_{a_2b_3}G^{(0)}_{a_3b_2} \nonumber \\ &&
- G^{(0)}_{a_1b_2} G^{(0)}_{a_2b_3}G^{(0)}_{a_3b_1}
- G^{(0)}_{a_1b_3} G^{(0)}_{a_2b_1}G^{(0)}_{a_3b_2}
+ G^{(0)}_{a_1b_3} G^{(0)}_{a_2b_2}G^{(0)}_{a_3b_1} ,
\end{eqnarray}
one also obtains $G^{{\rm
III}c}_{\sigma\sigma\sigma,\sigma\sigma\sigma}=0$. [Note that this
verification requires to consider 6!=360 different time sectors.] It
is clear that this result holds to all orders, i.e. 
\begin{equation}
G^{Rc}_{\sigma\cdots\sigma,\sigma\cdots\sigma}=0.
\end{equation}

The Green's function $G^{{\rm
II}c}_{\uparrow\downarrow,\uparrow\downarrow}$ can be calculated in the
same way: 
\begin{eqnarray}
G^{\rm
II}_{\uparrow\downarrow,\uparrow\downarrow}(\tau_1,\tau_2;\tau_3,\tau_4)
&=& e^{\mu(\tau_1-\tau_3)+(\mu-U)(\tau_2-\tau_4)}  \Bigl \lbrace
e^{-\beta\mu}\theta(\tau_1-\tau_2)\theta(\tau_2-\tau_4)\theta(\tau_4-\tau_3)
\nonumber \\ && 
-\theta(\tau_2-\tau_4)\theta(\tau_3-\tau_1)\Bigl[\theta(\tau_1-\tau_2)+\theta
(\tau_4-\tau_3)\Bigr]   
+e^{\beta(\mu-U)}\theta(\tau_4-\tau_3)\theta(\tau_3-\tau_1)\theta(\tau_1
-\tau_2) \Bigr \rbrace .
\label{GII1}
\end{eqnarray}
In Fourier space, we obtain from Eq.~(\ref{GII1}) the connected part
\begin{eqnarray}        
G^{{\rm II}c}_{\uparrow\downarrow,\uparrow\downarrow}
(i\omega_1,i\omega_2;i\omega_3(i\omega_4))  
&=& - G^{(0)}_\uparrow(i\omega_1) G^{(0)}_\downarrow(i\omega_2)
G^{(0)}_\uparrow(i\omega_3)
G^{(0)}_\downarrow(i\omega_4)(i\omega_2-i\omega_3-U) \nonumber  \\ && 
- \frac{1}{i\omega_1+i\omega_2+2\mu-U}\Bigl[ e^{-\beta\mu}
G^{(0)}_\uparrow(i\omega_1) G^{(0)}_\uparrow(i\omega_3) 
-e^{\beta(\mu-U)} G^{(0)}_\downarrow(i\omega_2)
G^{(0)}_\downarrow(i\omega_4) \Bigr] ,
\label{GII2}
\end{eqnarray}    
where $\omega_4=\omega_1+\omega_2-\omega_3$ is fixed by energy
conservation. $G^{{\rm II}c}$ is related to the two-particle vertex by 
\begin{equation}
G^{{\rm II}c}_{\uparrow\downarrow,\uparrow\downarrow}
(i\omega_1,i\omega_2;i\omega_3(i\omega_4))   
= - G^{(0)}_\uparrow(i\omega_1)G^{(0)}_\downarrow
(i\omega_2)\Gamma_{\uparrow\downarrow,\uparrow\downarrow}  
(i\omega_1,i\omega_2;i\omega_3,(i\omega_4))
G^{(0)}_\uparrow(i\omega_3)G^{(0)}_\downarrow(i\omega_4). 
\label{GaII}
\end{equation}
From Eq.~(\ref{GII2}), we then deduce 
\begin{equation}
\Gamma_{\uparrow\downarrow,\uparrow\downarrow}
(i\omega_1,i\omega_2;i\omega_3(i\omega_4)) =i\omega_2-i\omega_3-U
\end{equation} 
in the low-temperature limit, i.e by neglecting the second and third
terms of the rhs of Eq.~(\ref{GII2}). These terms correspond to the
contributions of the empty and doubly occupied sites, which, according
to the discussion of Sec.~\ref{sec:piat}, should be discarded. 

Let us now consider the effect of the Berry phase term 
\begin{equation}
S_{\rm at}-S^{(0)}_{\rm at} = \int d\tau A^0 f^*_\uparrow p^* f_\uparrow p 
\end{equation}
on the single particle Green's function $G$. The latter can be
calculated from
\begin{eqnarray} 
G_\uparrow (\tau) &=& - \frac{1}{Z^{(0)}_{\rm at}} \int d\lambda \int {\cal D}
[f,e,p,d] e^*(\tau)p(\tau)f_\uparrow(\tau) f^*_\uparrow(0)p^*(0)e(0) 
e^{-S_{\rm at}} , \nonumber \\
G_\downarrow (\tau) &=& - \frac{1}{Z^{(0)}_{\rm at}} \int d\lambda \int
{\cal D}  
[f,e,p,d] d(\tau)p^*(\tau)f_\downarrow(\tau)
f^*_\downarrow(0)p(0)d^*(0) e^{-S_{\rm at}} 
\end{eqnarray}   
by integrating explicitly the bosonic ($e,p,d$) and fermionic
($f_\sigma$) fields and the Lagrange multipliers. 
The calculation can be found in Appendix C of
Ref.~\onlinecite{ND}. Neglecting the Berry phase term, one recovers
the results obtained above [Eqs.~(\ref{Gat1})]. The first order
correction in $A^0=\langle{\bf\Omega}|\dot{\bf\Omega}\rangle$ is 
\begin{eqnarray}
G^{(1)}_{\uparrow} (\tau) &=& \int d\tau_1 \, 
G^{(0)}_{\uparrow} (\tau-\tau_1)A^0(\tau_1)G^{(0)}_{\uparrow}
(\tau_1) , \nonumber \\ 
G^{(1)}_{\downarrow} (\tau) &=& - \int d\tau_1 \, 
G^{(0)}_{\downarrow} (\tau-\tau_1)A^0(\tau_1)G^{(0)}_{\downarrow}
(\tau_1) 
\end{eqnarray}
in the limit $T\to 0$. 
These results are easily extrapolated to higher orders as (in matrix form)
\begin{eqnarray}
G_{\uparrow} &=& G^{(0)}_{\uparrow} + 
G^{(0)}_{\uparrow} A^0 G^{(0)}_{\uparrow} + 
G^{(0)}_{\uparrow} A^0 G^{(0)}_{\uparrow} A^0 G^{(0)}_{\uparrow} 
+ \cdots \nonumber \\ 
G_{\downarrow} &=& G^{(0)}_{\downarrow} - 
G^{(0)}_{\downarrow} A^0 G^{(0)}_{\downarrow} + 
G^{(0)}_{\downarrow} A^0 G^{(0)}_{\downarrow} A^0 G^{(0)}_{\downarrow} 
- \cdots 
\end{eqnarray}
or, equivalently,
\begin{eqnarray}
G^{-1}_{\uparrow} &=& G^{(0)-1}_{\uparrow} - A^0 , \nonumber \\ 
G^{-1}_{\downarrow} &=& G^{(0)-1}_{\downarrow} + A^0 .
\end{eqnarray}
Note that these Green's functions, like the gauge field $A^0$,
depend on the site which is considered.

\section{Closed loops in the strong-coupling perturbative expansion}
\label{appIV} 

In this Appendix, we discuss the vanishing of closed loops of
$\gamma_\uparrow$ particles in the strong-coupling perturbative expansion.  

Since the Green's functions $G^{Rc}$ are related to the vertices
$\Gamma^R$, a diagram can always be expressed in terms of the
$G^{Rc}$'s. Using Eqs.~(\ref{gamt}), one can show that $G^{Rc}$ is a sum of
terms proportional to $(-1)^P\prod_{i=1}^R
G_{\sigma_i}(\tau_i-\tau_{P(i)})$ where $P$ is a permutation of
$[1,R]$ satisfying $\sigma_{P(i)}=\sigma_i$ ($i\in [1,R]$). 
Each closed loop of $\gamma_\sigma$ particles will then give terms of the type
$G_\sigma(\tau_0-\tau_1)\hat t_{\sigma\sigma} \cdots \hat t_{\sigma\sigma}
G_\sigma(\tau_n-\tau_0)$ where $\sigma$ is the (conserved) spin along the
loop and the integer $n$ varies from $2$ to $\infty$. Terms
with $n=1$ correspond to anomalous diagrams and should be discarded. 
By summing over the different values of $n$, we obtain $\hat {\cal
G}^{(0)}_\sigma({\bf r},{\bf r};\tau=0^-)-G_\sigma({\bf r},{\bf
r};\tau=0^-)$ where $\hat {\cal 
G}^{(0)}_\sigma$ is defined by Eq.~(\ref{Gcal0}). In the absence of
particles in the UHB, both $\hat {\cal G}^{(0)}_\downarrow({\bf r},{\bf
r};\tau=0^-)$ and  $G_\downarrow({\bf r},{\bf r};\tau=0^-)$ are equal
to zero. We conclude that closed loops of $\gamma_\downarrow$
particles vanish. On the other hand, since  $\hat {\cal 
G}^{(0)}_\uparrow({\bf r},{\bf r};\tau=0^-)-G_\uparrow({\bf r},{\bf
r};\tau=0^-)=n-1$, closed 
loops of $\gamma_\uparrow$ particles do not vanish, except when the density of
particles $n$ equals 1 (half-filled case). 

\section{Green's function of the ``physical'' fermions}
\label{AppV}

In this section, we relate the ``physical'' Green's function $-\langle
c_ac^*_b\rangle$ to the propagator of the $\gamma$ field. Consider the
fermionic field $\phi$ defined in the spin reference frame determined
by ${\bf\Omega}$ [Eq.~(\ref{B1})]: 
\begin{equation}
c_{{\bf r}\sigma} = \sum_{\sigma'} (R_{\bf
r})_{\sigma\sigma'}\phi_{{\bf r}\sigma'} . 
\end{equation}
Since $\phi\equiv\gamma$ (in the space spanned by ${\cal
B}'_{{\bf\Omega}_{\bf r}}$), we deduce 
\begin{equation}
-\langle c_{{\bf r}\sigma}(\tau)c^*_{{\bf r}'\sigma}(\tau')\rangle= 
-\sum_{\sigma_1,\sigma_2} \langle (R_{\bf r}(\tau))_{\sigma\sigma_1}
\gamma_{{\bf r}\sigma_1}(\tau) \gamma^*_{{\bf r}'\sigma_2}(\tau')
(R^\dagger_{{\bf r}'}(\tau'))_{\sigma_2\sigma} \rangle .
\end{equation}
For fermions in the LHB, we then find
\begin{equation}
{\cal G}^{\rm LHB}_\sigma ({\bf r}\tau,{\bf r}'\tau') = 
-\langle (R_{\bf r}(\tau))_{\sigma\uparrow}
\gamma_{{\bf r}\uparrow}(\tau) \gamma^*_{{\bf r}'\uparrow}(\tau')
(R^\dagger_{{\bf r}'}(\tau'))_{\uparrow\sigma} \rangle .
\label{E1}
\end{equation}
Here $\gamma_\uparrow=e^*pf_\uparrow$,
$\gamma_\downarrow=p^*df_\downarrow$ and the mean values are to be
taken with the action $S[f,e,p,d,\lambda;{\bf\Omega}]$
[Eq.~(\ref{action1})].  The Green's function ${\cal G}^{\rm LHB}$ can
be expressed as the functional derivative of the partition function
calculated in the presence of external sources. It is then possible to
carry out the steps leading to the effective action
$S[\gamma^*,\gamma;{\bf\Omega}]$ [Eqs.~(\ref{action2}) and
(\ref{action3})]. Taking the functional
derivative, one finds that ${\cal
G}^{\rm LHB}$ is given by Eq.~(\ref{E1}) where the $\gamma$ field is
now an elementary field and the mean value has to be taken with the action 
$S[\gamma^*,\gamma;{\bf\Omega}]$. 

\eleq

{}

\ecols 


\begin{thebibliography}{}
\bibitem{note0.0} In particular, charge carrier motion and spin
fluctuations are essentially uncoupled at the mean-field
level. Presumably, a correct description of their mutual interaction
would require to go beyond mean-field theory.
\bibitem{note0} Here we use our notations rather than those of
Ref.~\onlinecite{Schulz90}.
\bibitem{note5} For a half-filled system, there
is no difference (as far as the free energy is concerned) between a
$t/U$ expansion and an expansion around the atomic limit (i.e. a $t$
expansion).
\end{thebibliography}
\end{document}